\documentclass[twocolumn,twocolappendix]{aastex63}
\usepackage{amsmath}
\usepackage{amssymb}
\usepackage{bm}
\usepackage{graphicx}
\usepackage{color}
\usepackage{float}

\definecolor{darkgreen}{rgb}{0,0.35,0}
\definecolor{blue}{rgb}{0,0,1}

\newcommand{\be}{\begin{eqnarray}}
\newcommand{\ee}{\end{eqnarray}}

\newcommand{\avg}[1]{\left\langle #1 \right\rangle}

\renewcommand{\vec}[1]{{\bm #1}}

\newcommand\as{\bgroup\markoverwith{\textcolor[rgb]{.5, 0, .6}{\rule[0.5ex]{8pt}{1.5pt}}}\ULon}

\newcommand\ys{\bgroup\markoverwith{\textcolor[rgb]{1.0, 0, 1.0}{\rule[0.5ex]{8pt}{1.5pt}}}\ULon}

\defcitealias{Li2021}{L21}

\begin{document}

\title{Hot Circumsingle Disks Drive Binary Black Hole Mergers in Active Galactic Nucleus Disks}

\author[0000-0002-7329-9344]{Ya-Ping Li}
\affiliation{Theoretical Division, Los Alamos National Laboratory, Los Alamos, NM 87545, USA}
\affiliation{Shanghai Astronomical Observatory, Chinese Academy of Sciences, Shanghai 200030, China}
\author[0000-0001-8291-2625]{Adam M. Dempsey}
\affiliation{Theoretical Division, Los Alamos National Laboratory, Los Alamos, NM 87545, USA}
\author[0000-0003-3556-6568]{Hui Li}
\affiliation{Theoretical Division, Los Alamos National Laboratory, Los Alamos, NM 87545, USA}
\author[0000-0002-4142-3080]{Shengtai Li}
\affiliation{Theoretical Division, Los Alamos National Laboratory, Los Alamos, NM 87545, USA}
\author[0000-0001-5550-7421]{Jiaru Li}
\affiliation{Theoretical Division, Los Alamos National Laboratory, Los Alamos, NM 87545, USA}
\affiliation{Center for Astrophysics and Planetary Science, Department of Astronomy, Cornell University, Ithaca, NY 14853, USA}

\correspondingauthor{Ya-Ping Li}
\email{liyp@shao.ac.cn}

\begin{abstract}

Binary black hole (BBH) mergers, particularly those with component masses in the pair-instability gap, may be produced by hierarchical mergers in the disks surrounding Active Galactic Nuclei (AGN).
While the interaction of an embedded BBH with an AGN disk is typically assumed to facilitate a merger, recent high-resolution hydrodynamical simulations challenge this assumption.
However, these simulations often have simplified treatments for the gas thermodynamics. 
In this work, we model the possible consequence of various feedback from an embedded BBH with a simple model that maintains an enhanced temperature profile around each binary component. 
We show that when the minidisks around each BH become hotter than the background by a factor of three, the BBH orbital evolution switches from expansion to contraction.
By analyzing the gravitational torque profile, we find that this change in direction is driven by a weakening of the minidisk spirals and their positive torque on the binary. 
Our results highlight the important role of thermodynamics around BBHs and its effect on their orbital evolution, suggesting that AGN disks could be efficient factories for BBH mergers. 
\end{abstract}

\section{Introduction}\label{sec:intro}

Active galactic nucleus (AGN) disks have been proposed as one of the most promising locations for producing some of the detected stellar-mass binary black hole (BBH) and neutron star mergers \citep[e.g.,][]{McKernan2012,Bartos2017,Stone2017,Leigh2018,Grobner2020,Kimura2021,Wang2021,Zhu2021}, especially for the recent detection of the heaviest BBH merger event GW190521 with a possible electromagnetic counterpart  \citep{Abbott2020b,Graham2020}.  This ``AGN channel" could be an intriguing alternative to other traditional formation channels, including isolated binary evolution\citep[e.g.,][]{Belczynski2016,Mandel2016}, triple and quadruple systems \citep[e.g.,][]{Fernandez2019,Fragione2019,Liu2021}, globular clusters \citep[e.g.,][]{Benacquista2013,Rodriguez2016}, and nuclear star clusters \citep[e.g.,][]{OLeary2009,Fragione2022}.

Most studies for binary mergers in AGN disks are  based on \textit{N}-body \citep[e.g.,][]{Secunda2019}, Monte Carlo \citep[e.g.,][]{Yang2019b,Tagawa2020,McKernan2021}, semianalytical estimates \citep{Samsing2020}, or hydrodynamical simulations \citep{Baruteau2011,Kaaz2021,LiLai2022}. 
Most of these works  assume that BBHs contract in AGN disks. 
However, this is not a certainty, as we have shown in our previous paper (\citealt{Li2021}, hereafter \citetalias{Li2021}) that prograde BBHs can expand their orbit when the circumsingle disks (CSD) around each BH are adequately resolved.
Past studies, such as \citet{Baruteau2011}, used excessively large gravitational softenings that effectively removed the important CSD spiral arms.
These spirals have been shown to be the key component in driving binary expansion, and so resolving them is crucial (e.g., \citealt{Tang2017,Moody2019,Duffell2020,Munoz2019,Tiede2020,Heath2020,DOrazio2021,Dittmann2021,Zrake2021}; \citetalias{Li2021}).

However, \citetalias{Li2021} assumed a smooth temperature over the whole disk, which may be unrealistic, especially for the regions within the Bondi/Hill radius of each BH.
In particular, more realistic treatments of the thermodynamics and viscosity are needed to take into account the feedback from the BH. 
Feedback may be driven by super-Eddington accretion onto stellar-mass black holes. 
Since luminous AGNs are known to accrete at a few percent of the Eddington rate for supermassive BHs (SMBH) \citep[e.g.,][]{Netzer2015}, if an embedded {\it stellar-mass} BH accretes at just a fraction of that rate, the resulting accretion rate will be orders of magnitude greater than its Eddington rate.

Numerical MHD simulations for super-Eddington accretion flows show that the midplane disk temperatures around stellar-mass black holes are usually around $\sim10^{7}\ {\rm K}$ \citep[e.g.,][]{Jiang2014}, which are 
expected to be much higher than those of the global disk due to higher mass BHs having lower disk temperatures \citep[e.g.,][]{Shakura1973,Abramowicz1980}. 

The heat released by the embedded luminous objects may impact how the binary orbit evolves over time \citep{Szulagyi2016,Masset2017,Hankla2020}. 
Additionally,  strong outflows and radiation from accretion disks around BHs could impose significant mechanical and radiative feedback effects on the surrounding environment \citep{Proga2002,Jiang2014,Yuan2018}. 
Numerous observational evidence for disk winds has also been accumulated both for cold and hot accretion flows around BHs \citep[e.g.,][]{Tombesi2015,Shi2021}. 
Those feedback effects could possibly change the thermodynamics around CSDs and thus modify the BBH evolution \citep{SouzaLima2017,delValle2018}.

In this work, we extend \citetalias{Li2021} to study the effect of CSD temperature structure on the BBH orbital evolution in AGN disks using 2D hydrodynamical simulations.  
To approximate the effect of feedback from each BH, we model the CSD temperature structure with a simple phenomenological model.

This \textit{Letter} is organized as follows. In Section~\ref{sec:method} we introduce the method and models presented in this work. 
In Section~\ref{sec:results} we explore the binary evolution for different CSD temperature profiles. 
Finally, we summarize the main results of this work and briefly discuss the implications for gravitational wave observations in Section~\ref{sec:conclusion}. 
\begin{figure*}[t]
\centering
\includegraphics[height=0.4\textwidth]{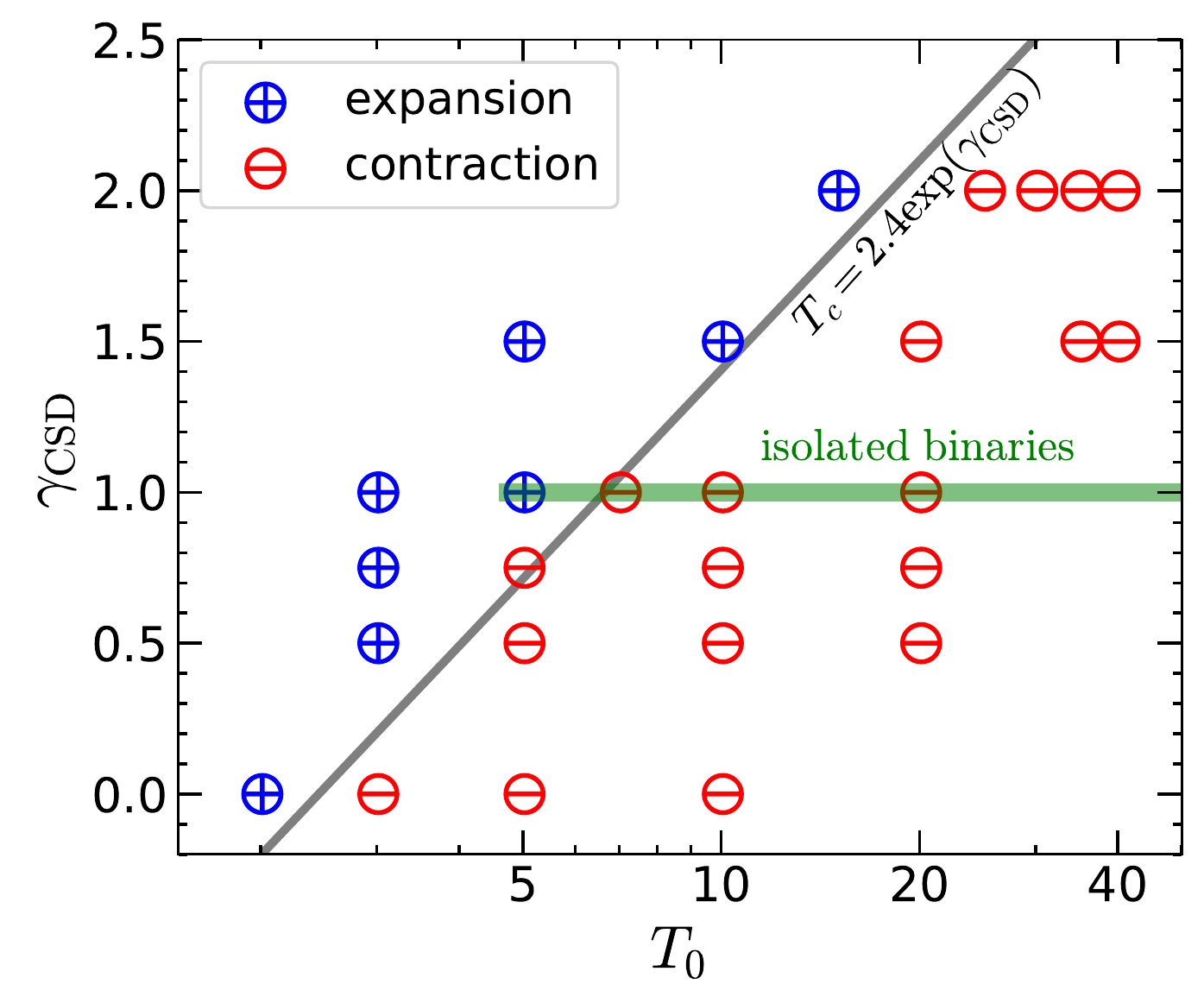} \\
\includegraphics[height=0.3\textwidth]{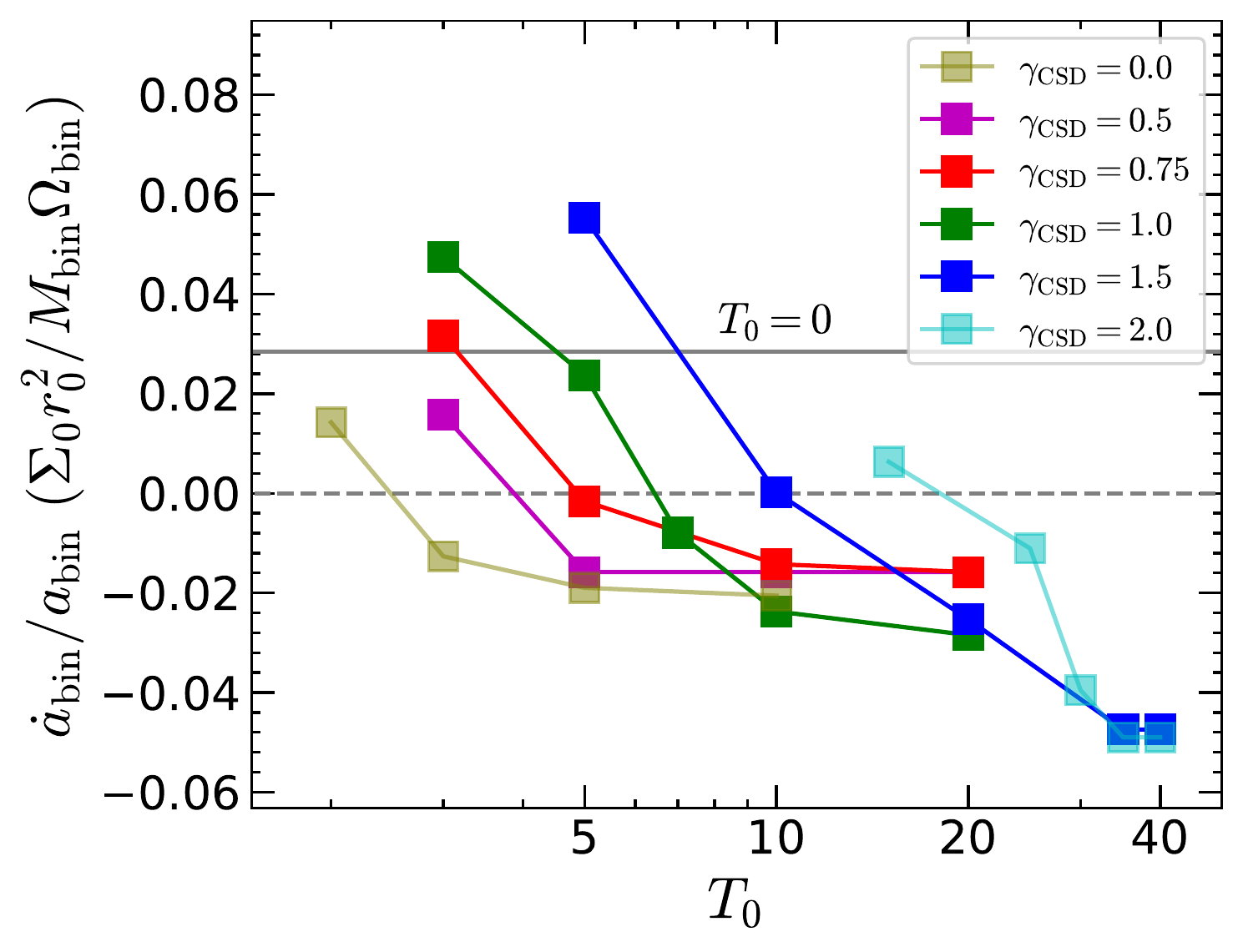}
\includegraphics[height=0.3\textwidth]{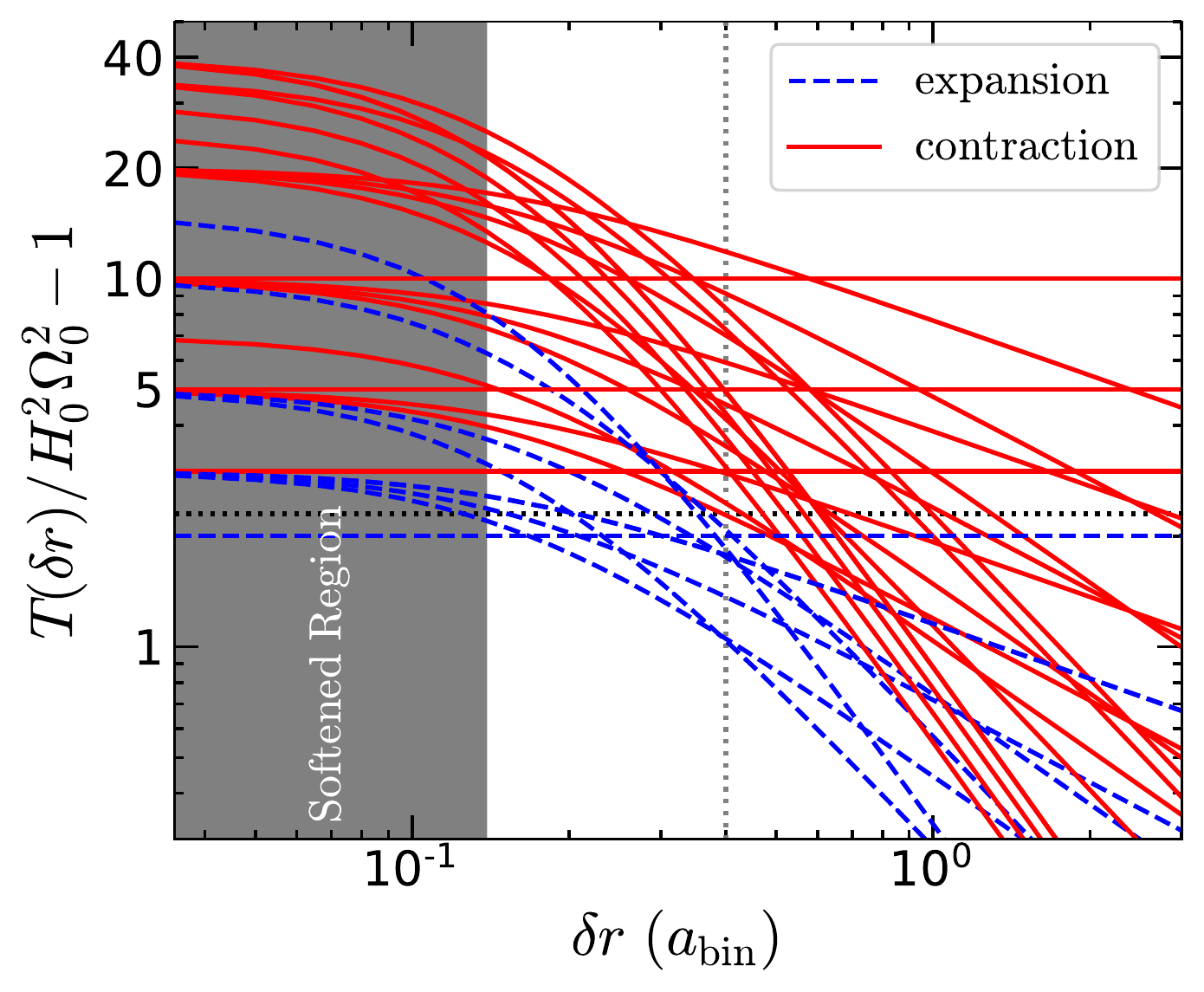}
\caption{
Upper panel: scatter plot of $\dot{a}_{\rm bin}/a_{\rm bin}$ for different $T_{0}$ and $\gamma_{\rm CSD}$. Symbols with plus ($\oplus$) and minus ($\ominus$) signs correspond to expanding and contracting BBHs. 
The gray line is the boundary between expansion and contraction, where $T_{\rm c} = 2.4 \exp(\gamma_{\rm CSD})$. 
The narrow green shaded region indicates the parameter space covered by isolated binary simulations ($\gamma_{\rm CSD}=1$), for which all models show that the BBHs are expanding \citep{Tang2017,Moody2019,Munoz2019,Munoz2020,Duffell2020,Tiede2020,Heath2020,DOrazio2021,Dittmann2021,Zrake2021}.  
Lower left panel: $\dot{a}_{\rm bin}/a_{\rm bin}$ as a function of $T_{0}$ for different $\gamma_{\rm CSD}$. The $\dot{a}_{\rm bin}/a_{\rm bin}$ of the control run with $T_{0}=0$ is shown as the black solid line. 
Lower right panel: the radial temperature profile around the CSD for all models. 
The blue dashed (red solid) lines correspond to expanding (contracting) BBHs. The vertical dotted line shows the size of CSD ($\delta r\approx0.4a_{\rm bin}$), and the horizontal dotted line illustrates the boundary between a ``hot" and ``cold" CSD where $T(0.4a_{\rm bin})/H_{0}^{2}\Omega_{0}^{2}\approx3.3$.
}
 \label{fig:adotall}
\end{figure*}

\begin{figure*}[t]
\centering
\includegraphics[width=0.95\textwidth,clip=true]{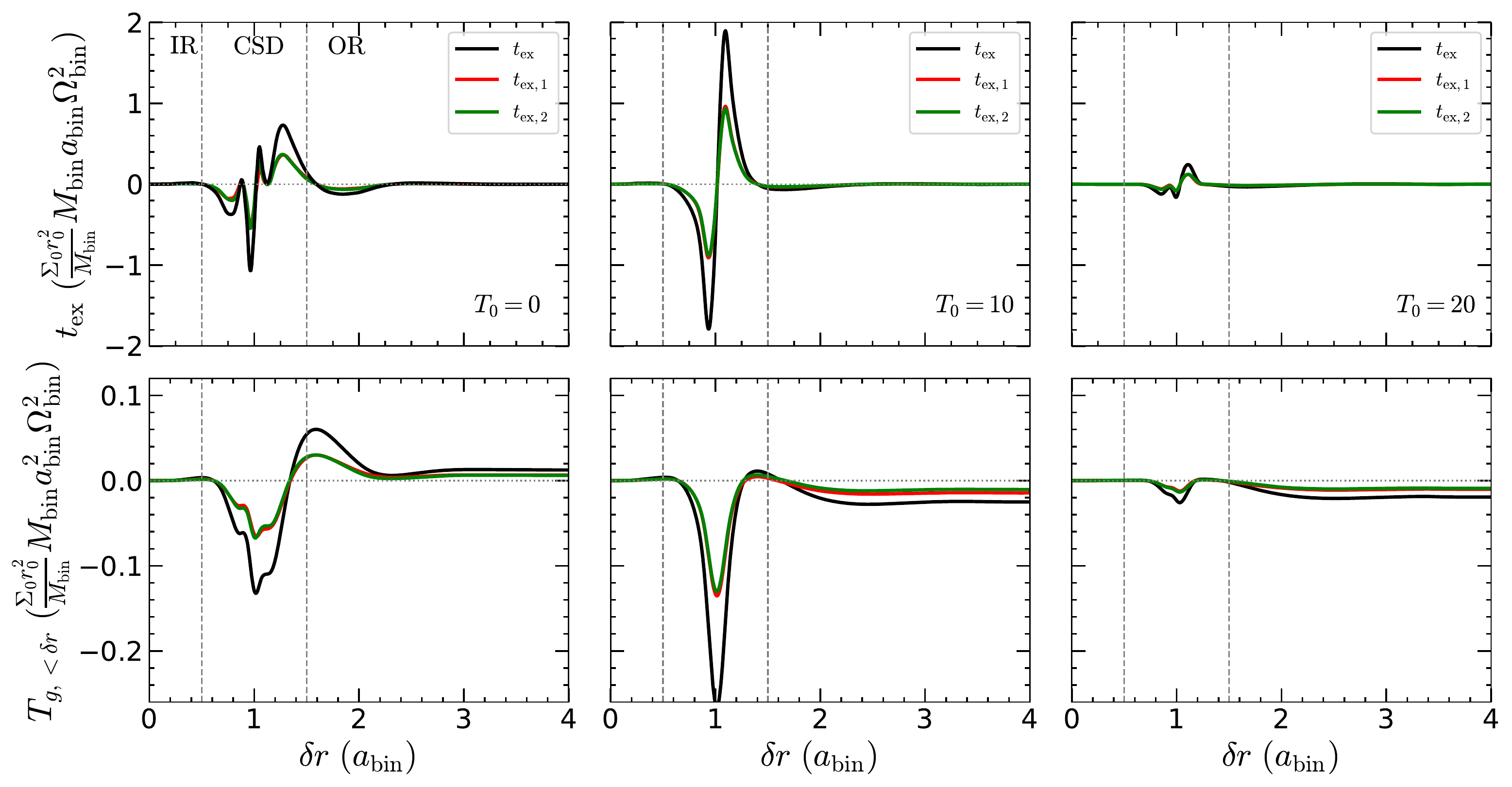}
\caption{
Azimuthally and time-averaged gravitational torques exerted on the binary component, which are calculated by centering on each BH for three simulations with $\gamma_{\rm CSD}=1$ and $T_0=0,10,$ and $20$. 
The upper panels display the torque density, $t_{\rm ex}$, while the lower panels show the cumulative torque $T_g = \int^{\delta r}_{0} t_{\rm ex}(r) dr$. 
Red and green lines correspond to the torques on each individual BH, while the black lines are their sum and correspond to the total torque on the BBH. 
The vertical dashed lines mark the three torquing regions discussed in the text.
Gas outside of $\delta r\sim 3\ a_{\rm bin}$ provides little to no torque on the BBH. 
Thus, the sign of $T_{\rm g}$ at that location determines whether the BBH contracts ($T_{\rm g} < 0$) or expands ($T_{\rm g} > 0$). 
}
 \label{fig:tex1d}
\end{figure*}

\begin{figure*}[t]
\centering
\includegraphics[width=0.98\textwidth,clip=true]{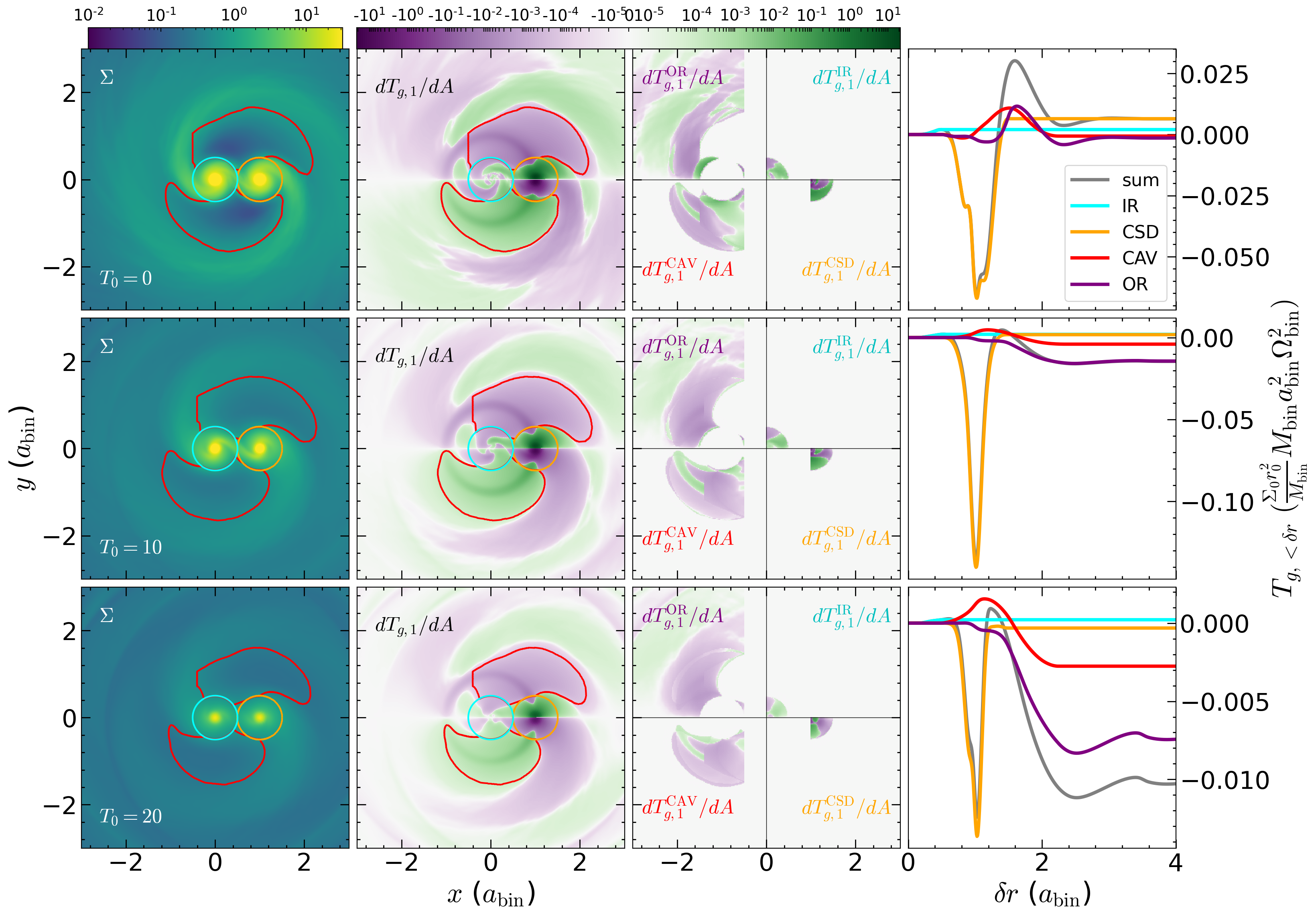}
\caption{Stacked gas surface density (first column) and specific gravitational torque density $dT_{\rm g,1}/dA$ on one component of the binary (second and third columns) for the simulations shown in Figure \ref{fig:tex1d}. The coordinate system $(x,y)$ is the frame corotating with BH 2. The summation of $dT_{\rm g,1}/dA$ (second column) over the symmetric center gives the symmetric component of the torque density (see Equation \ref{eq:texs}; third column).
The fourth column shows the cumulative radial torques for four different regions shown in the third column. 
Cold CSDs have stronger positive torques that can overcome the negative torques from the cavity and outer spiral arms.
}
 \label{fig:torqden}
\end{figure*}

\section{Method}\label{sec:method}

Most of the model setup and numerical methods are the same as \citetalias{Li2021}. 
Here we only briefly introduce the key features and defer a more detailed discussion to Appendix \ref{sec:app_method}.

We place an equal mass BBH on a prograde, circular orbit around a central SMBH.
The orbital radius and orbital frequency of the BBH's center of mass are denoted $r_0$ and $\Omega_0$, respectively. 
The BBH is initialized to a circular orbit with semi-major axis $a_{\rm bin} = 0.23 R_{\rm H}$, where $R_{\rm H} = [M_{\rm bin}/(3M_{\rm SMBH})]^{1/3} r_0$ is the Hill radius of the BBH, and we set $M_{\rm bin} = 0.002 M_{\rm SMBH}$. 

The AGN disk is modeled as an isothermal, 2D viscous  \citet{Shakura1973} accretion disk that we initialize to a vertically averaged surface density profile $\Sigma(r) \propto r^{-1/2}$ and a constant $\alpha$-viscosity with $\alpha=0.004$. 
We work in a cylindrical, $(r,\varphi)$, coordinate system centered on the SMBH with radial extent $[0.4,2.0] r_0$ and uniform resolution equaling $\sim 110$ cells within $R_{\rm H}$. 
The disk feels the gravitational force of the SMBH and BBH, the latter of which is softened with a cubic-spline potential \citep{Kley2009}. 
Unless otherwise stated, we set the gravitational softening length to $\epsilon_{\rm bh} = 0.15 a_{\rm bin}$. 
In contrast, the BBH does not feel the gravity of the disk.
Our results thus apply to the limit where the disk is low enough in mass that the time scale for the BBH orbit to change is slow compared to the time it takes for the disk to re-equilibrate.  

The disk boundary conditions are set to the initial conditions, and we apply wave-killing regions near both boundaries to prevent wave reflections  \citep{deValborro2006}.
Since we are only interested in the evolution of the BBH separation and not the center of mass motion, the large-scale flow near the boundaries is relatively unimportant. 
More sophisticated boundary conditions designed to obtain the correct gravitational torque on the center of mass of the BBH \citep[e.g.,][]{Dempsey2020} are thus not required. 

To model the effect of the heating from the BBH, we suppose that the gas cooling time scale is fast enough to maintain a temperature profile described by, 
\be \label{eq:temp}
\mathcal{R} T(\vec{r}) &=& H^2 \Omega^2  \left[1 + \sum_{i=1}^2 f(| \vec{r}-\vec{r}_{{\rm bh},i}|) \right],
\ee
where $\mathcal{R}$ is the specific gas constant (which we set to unity) and
\be 
f(\delta r) &=& T_0  \left( 1 +  \frac{\delta r^2}{\epsilon_{\rm bh}^2} \right)^{-\frac{\gamma_{\rm CSD}}{2}}  , 
\ee
is a function that specifies the local temperature enhancement around each BH. 
Far from the BBH, $\mathcal{R} T(r)\rightarrow H^2 \Omega^2$, where $H=0.05 r_0 (r/r_0)$ is the vertical disk scale height and $\Omega$ is the Keplerian frequency around the SMBH. 
The gas in the vicinity of each BH heats up to an axisymmetric profile following $\mathcal{R}T(r)\sim (1+T_0) H_0^2 \Omega_0^2/ \delta r_i^{\gamma_{\rm CSD}}$, where $\delta r_i = |\vec{r}-\vec{r}_{{\rm bh},i}|$. 
To maintain a finite temperature, we smooth the temperature enhancement to $T_0$ as the distance to the BH becomes less than the gravitational softening length. 
We additionally force $f(\delta r_i) =0$ when $\delta r_i \ge R_{\rm H}$ to maintain the background AGN disk temperature on large scales.

In Section~\ref{sec:results}, we use the hydrodynamical code \texttt{LA-COMPASS} \citep{Li2005,Li2009} to determine how the two-parameter temperature profile of Equation \eqref{eq:temp} modifies the torque of the disk on the BBH and the corresponding orbital evolution rate  $\dot{a}_{\rm bin}$. 
Previous studies have found that $\gamma_{\rm CSD}$ around accreting BHs is usually in the range of $\sim0.75-1.0$ \citep{Frank1992,Yuan2014,Jiang2014}. 
But, an even shallower profile could be attributed to a disk wind \citep{Li2019,Sun2019}. 
To cover a wide range of plausible values for $\gamma_{\rm CSD}$, we simulate disks with $\gamma_{\rm CSD}$ ranging from $0$ to $2$.

To keep our simulations relatively simple so as to isolate the effect of the temperature profile, we limit ourselves to studying 2D, locally isothermal, nonaccreting BBHs.
Our choice of BBH mass and disk scale height results in a value of $R_{\rm H} = 1.74 H$. 
Because of this, 3D effects may be subdominant and our 2D approximation is more realistic than a simulation with $R_{\rm H}<H$. 
Additionally, we neglect accretion onto the BHs to remove an additional parameter from our simulations (the mass removal rate).
Finally, our simple temperature model is meant to be a numerically clean approximation to BH feedback. 
In particular, we are implicitly assuming that any BH feedback is both localized to the BBH Hill sphere and is only in the form of a temperature enhancement around the binary. 
This simplification is a natural extension of our previous work, and it serves as a first step toward understanding more sophisticated radiation hydrodynamical simulations.

Each simulation is run until $\dot{a}_{\rm bin}$ reaches a quasi-steady state. 
We discuss the convergence of our simulations in Appendix~\ref{sec:app_conv} (see Figures \ref{fig:adot_time} and \ref{fig:mdot_time}).
Our choice of $M_{\rm bin}, H/r$, and $\alpha$ maintains the same gap profile as the more realistic parameters  $M_{\rm bin}=5\times10^{-5} M_{\rm SMBH}$, $H/r = 0.01$, and $\alpha=0.01$ (\citealt{Lin1993}; \citealt{Baruteau2011}; \citetalias{Li2021}).

\section{Results}\label{sec:results}

\subsection{Contracting Binaries Have Hot CSDs}

Our main conclusion is shown in the first panel of Figure \ref{fig:adotall} which plots the sign of $\dot{a}_{\rm bin}$ in the parameter space of $(T_0,\gamma_{\rm CSD})$ for all of our simulations. 
For each $\gamma_{\rm CSD}$, we identify the value of $T_0$ at which the BBH switches from expansion to contraction. 
We find that ``cold" CSDs result in binary expansion, whereas ``hot" CSDs result in binary contraction. 
The second panel of Figure \ref{fig:adotall} displays the actual steady-state values of $\dot{a}_{\rm bin}$ for each $\gamma_{\rm CSD}$ set as a function of their $T_0$ values. 
Noticeably, the magnitude of $\dot{a}_{\rm bin}$ saturates as $T_0$ increases for each value of $\gamma_{\rm CSD}$. 
There is a clear boundary separating the hot, contracting binaries from the cold, expanding binaries that lie along the relation $T_0 \approx 2.4 \exp(\gamma_{\rm CSD})$ (shown as the gray line in the first panel). 

In the third panel of Figure \ref{fig:adotall}, we show that this critical $T_0$ neatly corresponds to a minimum CSD temperature. 
We plot the radial CSD temperature profiles from Equation \eqref{eq:temp} for all of our models and color each line blue for expansion and red for contraction. 
The radial extent of the CSD is $\approx 0.4 a_{\rm bin}$ \citep{Artymowicz.1994}, and we see that at that distance, all of the contracting binaries have a CSD temperature greater than a factor of $\sim 3$ above the background. 
Thus, when we refer to a ``hot" CSD, we are referring to the \textit{entire} CSD being hotter than the background AGN disk by a factor of $\sim 3$. 
``Cold" CSDs can still have hot cores, but a steep temperature profile keeps the outer CSD below this threshold and the binary expanding.

Most importantly, we find that even a temperature enhancement as low as $T_{0}\gtrsim5$ for a shallow temperature profile (i.e., $\gamma_{\rm CSD}\lesssim0.75$) can alter the BBHs from expansion to contraction. Such a small $T_{0}$ could be easily satisfied in a realistic accretion disk environment as discussed in Section~\ref{sec:intro}. 
This indicates the inspiral of BBHs in AGN disks may be quite common after considering the hot CSD structures. 
This also suggests the importance of correct thermodynamics of CSDs around the BBH to account for the specific role of AGN disks in BBH merger events.

It should be noted that the softened regions around each BH, which are not modeled accurately, could modify the $\dot{a}_{\rm bin}$ values in Figure~\ref{fig:adotall}.  
We have checked that if we remove the contribution to $\dot{a}_{\rm bin}$ from within $0.5 \epsilon_{\rm bh}$\footnote{We only exclude the torque within $0.5\epsilon_{\rm bh}$ regions since our spline softening does not significantly modify the rotation velocity around each BH until within the region $0.5\epsilon_{\rm bh}$. }, the resulting $\dot{a}_{\rm bin}$ values are approximately the same as those shown in Figure \ref{fig:adotall}. 
More importantly, the boundary between expansion and contraction remains the same. 
In fact, the thickness of the gray line in the first panel reflects the change in the boundary when we include or exclude the softening region from the calculation of $\dot{a}_{\rm bin}$.

\subsection{Hot CSDs Have Weaker Torques}\label{sec:torq_decomp}
To understand why a BBH with a hot CSD contracts, we analyze the BH-disk torque\footnote{Technically it is the change in the orbital energy of the binary, not the torque, which determines $\dot{a}_{\rm bin}$ (see Appendix~\ref{sec:app_conv} and Equation \ref{eq:dotab}). But, for circular, nonaccreting binaries, the two can be interchanged with a small error of order $\partial_t (e_b^2)$.}
 distribution in this section.
We focus on three typical models with $\gamma_{\rm CSD}=1$ and $T_0$ values of $0,10,$ and $20$. 

To derive the torque profiles, consider the specific angular momentum of the binary $\ell_{\rm bin} = \vec{r}_{\rm bin} \times \dot{\vec{r}}_{\rm bin}$, where $\vec{r}_{\rm bin} = \vec{r}_{{\rm bh},1} - \vec{r}_{{\rm bh},2}$ is the BBH separation vector. 
The binary acceleration is made up of three terms, 
\be
 \ddot{\vec{r}}_{\rm bin} = - \frac{ G M_{\rm bin}}{|r_b|^3} \vec{r}_{\rm bin} + \delta \vec{f}_{\rm smbh} + \delta \vec{f}_{\rm disk} .
\ee
These are the Keplerian, SMBH, and disk accelerations, respectively. 
The first two are taken into account in our \texttt{LA-COMPASS} simulations, whereas the disk acceleration, $\delta \vec{f}_{\rm disk} = \vec{f}_{\rm grav,1} - \vec{f}_{\rm grav,2}$ is measured from the simulation output. 
A nonzero disk force torques the binary, changing its $\ell_{\rm bin}$ at the rate $\dot{\ell}_{\rm bin} = \vec{r}_{\rm bin} \times \delta \vec{f}_{\rm disk}$. 
This expression can be rewritten  as,
\be
\dot{\ell}_{\rm bin} = \left( (\vec{r} - \vec{r}_{\rm bh,2} ) - (\vec{r} - \vec{r}_{\rm bh,1}) \right) \times (\vec{f}_{\rm grav,1} - \vec{f}_{\rm grav,2}) ,
\ee
and because $\vec{f}_{{\rm grav},i} \propto (\vec{r}-\vec{r}_{{\rm bh},i})$, $\dot{\ell}_{\rm bin}$ simplifies to only two terms, 
\be \label{eq:torque_1}
\dot{\ell}_{\rm bin}  =  (\vec{r} - \vec{r}_{\rm bh,2} ) \times \vec{f}_{\rm grav,1} + (\vec{r} - \vec{r}_{\rm bh,1} ) \times \vec{f}_{\rm grav,2} .
\ee
The first term can be thought of as the torque on BH 1 measured in a coordinate system centered on BH 2, and vice-versa for the second term. 
In other words, Equation \eqref{eq:torque_1} can be written in the more familiar form,

\begin{eqnarray}
    \dot{\ell}_{\rm bin} &=& \sum_{j=1}^{2} \int \Sigma \frac{\partial \Phi_{{\rm bh},j}}{\partial \varphi_i} \delta r_i d\delta r_i d\varphi_i 
    \label{eq:dotlb2}
\end{eqnarray}
where $\delta r_i \equiv | \vec{r} -\vec{r}_{{\rm bh},i}|$ and $\varphi_i$ define a cylindrical coordinate system centered on BH $i$, and $\Phi_{{\rm bh},j}$ is the potential of the companion BH $j$.
As such, we define two torque densities, 
\begin{equation}
dT_{{\rm g},1}/dA=\Sigma_{2}^{\prime}\partial\Phi_{\rm bh,1}/\partial\varphi_{2}
\end{equation}
and  
\begin{equation}
dT_{{\rm g},2}/dA=\Sigma_{1}^{\prime}\partial\Phi_{\rm bh,2}/\partial\varphi_{1},
\end{equation}
where $\Sigma_{i}^{\prime}$ are the nonaxisymmetric surface densities around each BH and $dA$ is the area element for the appropriate cylindrical coordinate system.
The total torque density for the BBH is then $\dot{\ell}_{\rm bin}=\int (dT_{{\rm g},1}/dA+dT_{{\rm g},2}/dA) dA$.

Figure~\ref{fig:tex1d} shows the azimuthally and time-averaged specific torques $t_{{\rm ex},i}= \delta r_j \int dT_{{\rm g},i}/dA d\varphi_j$ (upper panels), and cumulative torques $T_{{\rm g},i} = \int_0^{\delta r_j} t_{{\rm ex},i}(r) dr$ (lower panels), acting on  BH $i$ centered on BH $j$.
These are time-averaged over $600$ BBH orbits once the simulation has reached a steady state. 
Each column focus on one value of $T_0$, and for each panel we show both the total (black lines), as well as the individual torques on each BH (green and red lines).
The total torque on the binary is the constant value of $T_{\rm g,1}+T_{\rm g,2}$ at $\delta r \gg 3a_{\rm bin}$. 
This value is positive for $T_0=0$ (indicating expansion), and negative for $T_0=10$ and $20$ (indicating contraction). 
Note that all of the radial profiles are symmetric with respect to each BH. 
We thus focus only on the torque distribution acting on one of the BHs.

We divide the domain into three important regions indicated by the vertical lines.
The inner region (IR; $\delta r\le 0.5\ a_{\rm bin}$) is the torque on the BH from the companion BH's CSD.
This region tends to torque the binary up ($t_{\rm ex} > 0$), but is weak overall. 
Conversely, the outer region (OR; $\delta r\ge 1.5\ a_{\rm bin}$) tries to torque the binary down ($t_{\rm ex} < 0$) and is an important contribution to the overall torque. 
Lastly, there is the CSD region ($0.5\ a_{\rm bin} < \delta r < 1.5\ a_{\rm bin}$). 
Gas in this region experiences the strongest torque from the BH, and the total torque from the CSD depends on how asymmetric the torque from the inner CSD region (where $t_{\rm ex} <0$) compares to the torque from the outer CSD region (where $t_{\rm ex}>0$).
We see that in most cases, the outer CSD contributes slightly more torque leading to a net positive torque coming from the BH's CSD. 
The magnitude of the CSD torque, however, becomes weaker overall as the CSD temperature increases.
In fact, at $T_0=10$, the positive CSD torque is so weak that the negative OR torque results in the binary contracting. 

Quantitatively, the torque from the CSD region, normalized to $\Sigma_{0}r_{0}^{2} a_{\rm bin}^2 \Omega_{\rm bin}^2$, goes from $5.0\times10^{-2}$ when $T_0=0$ to $5.0\times10^{-3}$ when $T_{0}=10$, and even to negative $-2.5\times10^{-3}$ when $T_0=20$. 
On the other hand, the torque from the OR region only falls from $-4.2\times10^{-2}$ when $T_0=0$ to  $-1.8\times10^{-2}$ when $T_0=20$. 
As the CSDs get hotter, the magnitude of $t_{\rm tex}$ decreases overall.

We provide more context for the three regions we have just described in Figure \ref{fig:torqden} where we show the 2D, time-averaged distributions of surface density and torque density. 
These profiles are calculated from stacking a large number of snapshots in the rotating frame of the binary (\citealt{Munoz2019}; \citetalias{Li2021}), centered on one of the BHs and plotted in Cartesian $(x,y)$ coordinates. 
From the surface density in the first column and the torque density in the second column we divide the space up into four regions shown by the contours. 
There are the IR and CSD regions defined by the cyan and orange circles, respectively, with radii $0.5 a_b$ centered on each BH. 
Additionally, there is the new cavity region (denoted CAV) that lies within the red contours. 
This is meant to approximate the location of the low-density regions between the BHs' CSDs and the outer spiral arms. 
Finally, everything not enclosed by a contour we place into the OR region. 
The fourth column shows the cumulative torque profiles for each of these four regions.

From the second column, we see that the torque density for each region has a large antisymmetric component about $y=0$ and $x=x_c$, where $x_c=0$ for the IR region, $x_c=0.5 a_{\rm bin}$ for the CAV and OR regions, and  $x_c = a_{\rm bin}$ for the CSD region. 
The antisymmetric component does not contribute to the total torque (i.e., the green subregions mostly cancel the purple subregions), and so in the third column we show only the symmetric component of the torque for each region. 
To do this we define the two symmetric torque densities (one about each axis), 
\be
\left. \frac{dT_{g,i}}{dA} \right|^{x}  &=& \frac{d T_{g,i}}{dA}( +|x-x_c|, y) + \frac{d T_{g,i}}{dA}( -|x-x_c|,y) , \nonumber \\
\left. \frac{dT_{g,i}}{dA} \right|^{\rm sym} &=& \left.\frac{d T_{g,i}}{dA}\right|^x ( x, +|y|) + \left.\frac{d T_{g,i}}{dA}\right|^x(x, -|y|) , \label{eq:texs}
\ee

and then plot the $dT_{g,i}/dA|^{\rm sym}$ for each region in one of the four quadrants in the third column.

Using the symmetric torque and $T_g$ profiles, we come to the following picture of the disk torque on the binary and its dependence on temperature. 
There is a competition between the positive torques from the IR $+$ CSD regions and the negative torques from the CAV $+$ OR regions. 
At low temperature enhancements (e.g., $T_0=0$), there is generally more positive torque coming from each BH's CSD that can overcome the CAV and OR torques. 
When this happens, the relative contribution between the IR and CSD is typically in favor of the CSD. 
In the $T_0=0$ example, the CSD contributes roughly three times as much positive torque as the IR region.

As $T_0$ increases, two things happen. 
First, the spiral arms in the CSDs weaken and their corresponding positive torque decreases. 
There is roughly a factor of two drop in the IR $+$ CSD torque going from $T_0=0$ to $T_0=10$. 
Second, the torque from the OR region increases dramatically -- almost by a factor of $10$ across the same $T_0$ jump. 
The torque from the CAV region also increases, but less dramatically (a factor of $\sim 6$ increase). 
These two effects combined result in a net negative torque on the BBH in the $T_0=10$ case. 
Going to $T_0=20$ almost completely removes the torque from the BHs' CSDs and only slightly decreases the total torque from the CAV $+$ OR regions.

As a final point, we note that it is \textit{not} the relative temperatures between the CSDs and circumbinary disks (CBDs) that determine whether the binary expands or contracts. 
Rather, it is the overall magnitude of the CSD and CBD temperatures that matters. 
This can be seen by our results at $\gamma_{\rm CSD}=0$, which raise the CSD and CBD temperatures by the same uniform factor. The fact that these simulations also find that BBHs contract when the CSD is hotter than the background by a factor of $\sim 3$, indicates that the CSD torques weaken enough for the negative CBD torque to set the overall sign of the torque.

\section{Conclusions and Discussion}\label{sec:conclusion}

We have performed a series of global 2D hydrodynamical simulations of an equal mass, circular  BBH embedded in an accretion disk to study the effect of the CSD temperature on the evolution of the binary separation. 
The CSD temperature profile around each BH is approximated with a phenomenological power-law model designed to mimic the feedback from each BH.  
We find that when the entire CSD is a factor of $\sim 3$ hotter than the AGN disk, the positive CSD torques are suppressed enough so that the negative outer torques drive the binary to an eventual merger.

A hot CSD can likely arise from strong feedback fueled by super-Eddington accretion onto the BH.
This feedback could take the form of jets, winds, and strong radiation \citep[e.g.,][]{Jiang2014,Wang2021}.
All of these can modify the flow structures, e.g. whether the CSDs can exist or not, in addition to their effect on the thermodynamics around BBHs. We should point out that we do not consider the accretion onto the BBH in the current work. 
Although \citetalias{Li2021} suggests that accretion has a minor effect on the binary orbital evolution for cold CSDs, its effect when coupled with the modified thermodynamics around BBHs still needs to be addressed in the future with detailed simulations.

It is worth noting that almost all simulations of isolated binaries use a temperature profile equal to $\mathcal{R} T(\vec{r}) \approx h_0^2 \sum_i GM_i/|\vec{r}-\vec{r}_{{\rm bh},i}|$, which is similar to Equation \eqref{eq:temp} with $\gamma_{\rm CSD}=1$ \citep[e.g.,][]{Tang2017,Moody2019,Munoz2019,Munoz2020,Duffell2020,Tiede2020,Heath2020,DOrazio2021,Dittmann2021,Zrake2021}.
The green shaded region in Figure \ref{fig:adotall} shows the large range of $T_0$ values used in a sample of those studies.
The temperature enhancement in the CSDs can get up to well over a factor of $40$ in some cases.

Despite the fact that the CSDs found in isolated binary simulations are ``hot", most of these studies still find binary expansion\footnote{Notable exceptions that find contracting binaries are when the binary is sufficiently eccentric \citep{Munoz2019,DOrazio2021}; when the binary mass ratio is sufficiently small \citep{Duffell2020,Dempsey2021}; or when the disk is sufficiently cold \citep{Tiede2020,Heath2020,Dittmann2022}.}.
We suspect that this discrepancy lies in the strength of the cavity and outer spiral torques for an isolated binary.
In particular, simulations of isolated binaries find much wider and deeper cavities than embedded binaries. 
This makes the negative cavity and outer region torques on an embedded binary much stronger compared to an isolated binary simply because these regions are ``closer" to the binary (i.e., the cavity is shallower and filled in more).
This means that in order for an embedded binary to expand, it needs a much stronger CSD torque compared to an isolated binary. 

If this picture is correct, the transition point in $T_0$ for an isolated binary should be at a much larger value than what we find for an embedded binary. 
This also explains why simulations of isolated binaries in colder disks find contraction \citep{Tiede2020,Heath2020,Dittmann2022}.
These studies find that lowering the disk temperature (at fixed viscosity) raises the density near the edge of the cavity, i.e., the peak of the azimuthally averaged surface density profile in the CBD moves closer to the binary (see Figure 10 in \citealt{Tiede2020} and Figure 7 in \citealt{Dittmann2022}).
Because the cavity wall is closer to the binary and at a higher density, this strengthens the negative torques from this region, making it harder for the binary's CSD torques to push it apart.
Indeed, Figure 8 in \citet{Tiede2020} shows that the ``cavity" torque becomes more negative as the disk becomes colder.

Our study highlights the important role of thermodynamics around BBHs in their orbital evolution. 
Because our temperature model is only phenomenological, further work is required to self-consistently determine the CSD temperature profile.
In particular, this will require simulations with treatments for radiation and possible feedback from the BH. 
Calculating the correct feedback mechanism will require simulations near the actual accretion region of the BH which can only be captured by detailed general-relativistic radiative magnetohydrodynamic (GRRMHD) simulations. 
Such results would then need to be fed into the large-scale simulations presented here -- possibly with a subgrid-scale model.
The effects of magnetic fields and disk self-gravity may also be important for the structure of the CSDs. 
Finally, these effects should be included in fully 3D simulations to capture the full BBH-disk interaction problem.

\acknowledgments
We thank very beneficial comments from Zoltan Haiman and a very constructive report from the referee that helps to clarify some statements of the paper. We gratefully acknowledge the support by LANL/LDRD under project
number 20220087DR. 
This research used resources provided by the Los Alamos National Laboratory Institutional Computing Program, which is supported by the U.S. Department of Energy National Nuclear Security Administration under Contract No. 89233218CNA000001. 
This work is approved for unlimited release under LA-UR-21-31722.
Softwares: \texttt{LA-COMPASS} \citep{Li2005,Li2009}, \texttt{Numpy} \citep{vanderWalt2011}, \texttt{Scipy} \citep{Virtanen2020}, \texttt{Matplotlib} \citep{Hunter2007}.

\begin{appendix}

\section{Numerical Method}\label{sec:app_method}

We briefly describe the equations solved in our hydrodynamical simulations. We numerically solve the continuity and momentum equation for the gaseous disk with an embedded BBH, 
\be
\frac{\partial \Sigma}{\partial t} &+& \nabla \cdot (\Sigma \vec{v}) =0, \label{eq:mass_cons} \\
\frac{\partial (\Sigma\vec{v})}{\partial t} &+& \nabla\cdot(\Sigma \vec{v}\vec{v}) =-\Sigma\nabla\Phi-\nabla P+\Sigma \vec{f}_{\nu}, \label{eq:momen_cons}
\ee
where $\vec{v}$ is the gas fluid velocity, $\vec{f_{\nu}}$ is the viscous force from the \citet{Shakura1973} disk , and $P$ is the gas pressure determined from Equation~\eqref{eq:temp} with an isothermal equation of state. The total gravitational potential of the SMBH-BBH system is
\begin{eqnarray}
    \Phi &=& -\frac{G M_{\rm SMBH}}{\left|\bm{r}\right|} \nonumber \\
    &+& \sum_{i=1}^{2} \frac{M_{{\rm bh},i}}{M_{\rm SMBH}}\left[-\frac{G M_{\rm SMBH}}{D_{i}} +  \Omega_{{\rm bh},i}^{2} \bm{r}_{{\rm bh},i} \cdot \bm{r}\right],
    \label{eq:potential}
\end{eqnarray}
where $\Omega_{{\rm bh},i}$ is the orbital frequency of each BH with respect to the SMBH, and 
\begin{equation}
D_i^{-1}(\delta r_i) =
\begin{cases}
      \frac{1}{\delta r_{i}}\left[\left(\frac{\delta r_{i}}{\epsilon_{\rm bh}}\right)^{4}-2\left(\frac{\delta r_{i}}{\epsilon_{\rm bh}}\right)^{3}+2\frac{\delta r_{i}}{\epsilon_{\rm bh}}\right], & \text{if} \ \delta r_{i} \leqslant \epsilon_{\rm bh} \\
      \frac{1}{\delta r_{i}}, & \text{if} \ \delta r_{i} > \epsilon_{\rm bh}
\end{cases} 
\label{eq:sft}
\end{equation}
determines the softened potential from each BH \citep{Kley2009}. The terms in the bracket of Equation~\eqref{eq:potential} describe the potential from the BBH in Equation~\eqref{eq:dotlb2}.

\section{Convergence}\label{sec:app_conv}

\begin{figure}[htbp]
\centering
\includegraphics[width=0.45\textwidth,clip=true]{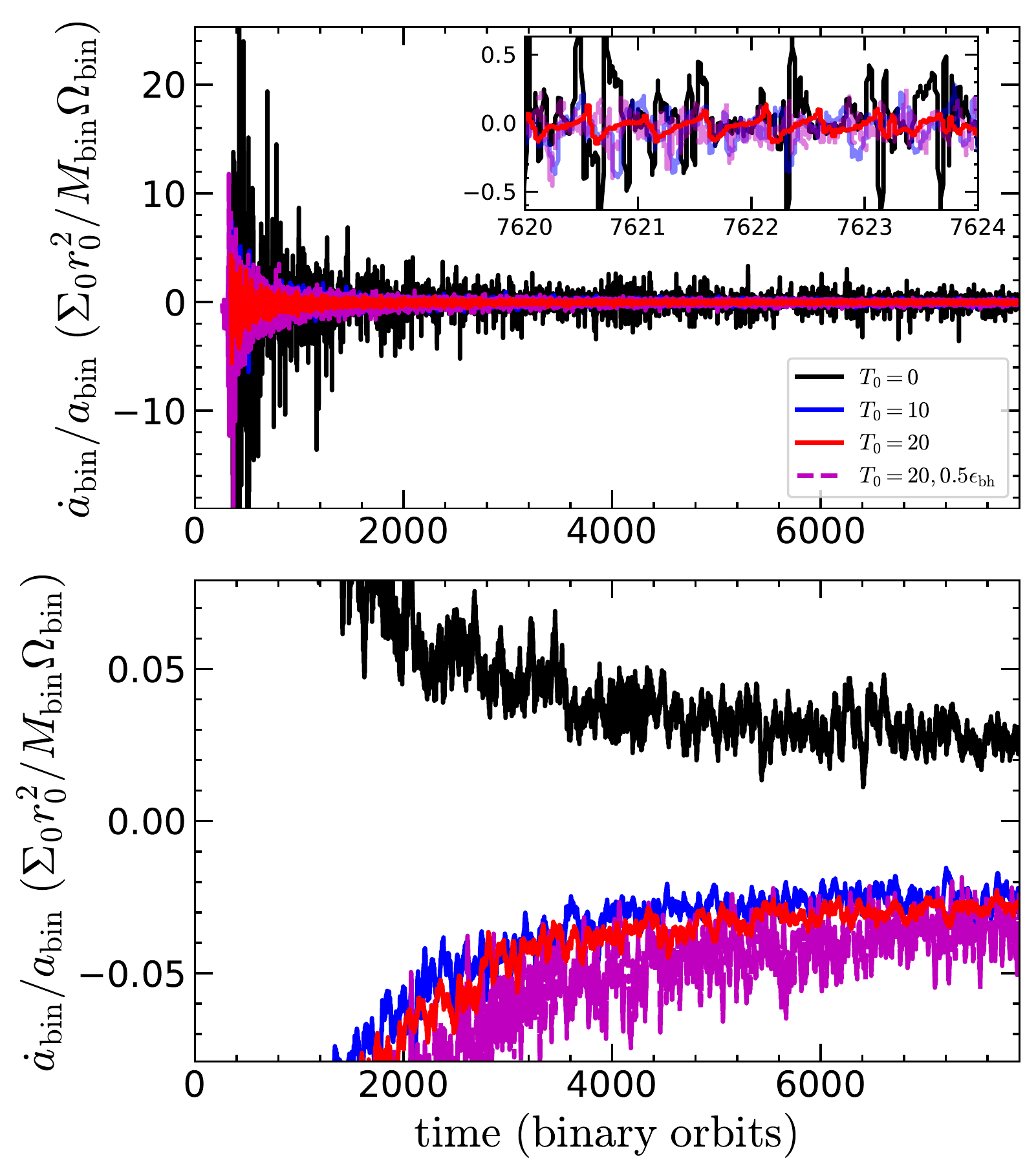}
\caption{Binary semi-major axis evolution rate for four models with $T_{0}=0$ (black lines), $T_{0}=10$ (blue lines), $T_{0}=20$ (red lines), and $T_{0}=20$ with a smaller softening of $0.075a_{\rm bin}$ (magenta dashed lines). 
All of the models with $T_{0}>0$ adopt $\gamma_{\rm CSD}=1.0$.  The time is measured in units of the BBH orbital period. The upper panel shows the $\dot{a}_{\rm bin}/a_{\rm bin}$ without time-averaging, and the lower panel is time-averaged over about two binary orbits.
All simulations converge in approximately $\sim 4000$ BBH orbits.
}
 \label{fig:adot_time}
\end{figure}

In this section, we show that our simulations are in steady state by verifying that $\dot{a}_{\rm bin}$ is converged in time, and that the radial mass accretion profile centered on the BBH is spatially constant. 
The binary semi-major axis evolution rate $\dot{a}_{\rm bin}/a_{\rm bin}$ due to the gravitational force from the gaseous disk is \begin{equation}
    \frac{\dot{a}_{\rm bin}}{a_{\rm bin}}=-\frac{\dot{\varepsilon}_{\rm bin}}{\varepsilon_{\rm bin}},
\label{eq:dotab}
\end{equation}
where  $\varepsilon_{\rm bin} = - G M_{\rm bin}/(2 a_{\rm bin})$ is the specific energy of the binary. 
We measure the power delivered to the binary by the disk as,
\begin{equation}
    \dot{\varepsilon}_{\rm bin} =(\dot{\bm{r}}_{{\rm bh},1}-\dot{\bm{r}}_{{\rm bh},2})\cdot(\bm{f}_{\rm grav,1}-\bm{f}_{\rm grav,2}).
\label{eq:dotepb}
\end{equation}
This is used to compute $\dot{a}_{\rm bin}$ every time-step of the simulation.

\begin{figure}[htbp]
\centering
\includegraphics[width=0.45\textwidth,clip=true]{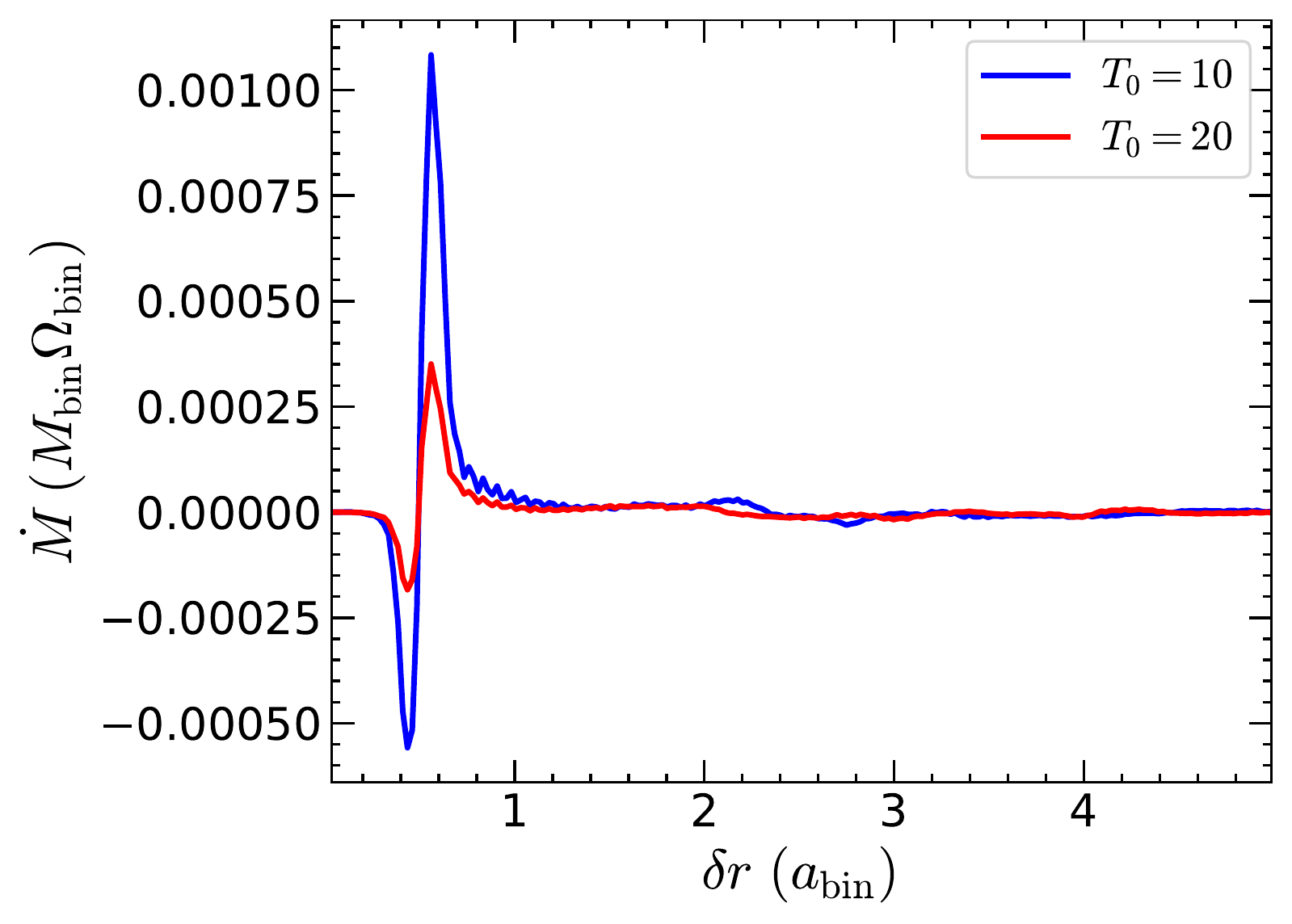}
\caption{Azimuthally and time-averaged accretion rates across the CBD for model $T_{0}=10$ (blue line) and $T_{0}=20$ (red line) at the end of our simulations. Both are with $\gamma_{\rm CSD}=1.0$. The accretion rates are calculated in the frame comoving with the center of mass of the BBH (not relative to each BBH component as defined previously.) A nearly constant zero rate far from the BBH suggests a steady state for the nonaccreting BBH.
}
 \label{fig:mdot_time}
\end{figure}

The time evolution of $\dot{a}_{\rm bin}/a_{\rm bin}$ for four models is shown in Figure~\ref{fig:adot_time}. The black line is the model without any enhancement of CSD temperature compared to the global disk (i.e., $T_{0}=0$), while the blue and red lines correspond to the case of $T_{0}=10$ and $T_{0}=20$, respectively. 
The magenta lines show the model with $T_{0}=20$ but decreasing the softening scale to $0.075a_{\rm bin}$ and increasing the resolution by a factor of two both radially and azimuthally ($n_{r}\times n_{\phi}=4096\times16384$).
Although the raw data without time-averaging in the upper panel shows strong variabilities, the running window time-averaging plot in the lower panel shows that all $\dot{a}_{\rm bin}/a_{\rm bin}$ values reach a steady state after $\sim 4000$ BBH orbits.

As we decrease the softening scale by half, the $\dot{a}_{\rm bin}/a_{\rm bin}$ evolution does not change significantly compared with the two $T_{0}=20$ models shown in Figure~\ref{fig:adot_time}, except that the oscillation becomes stronger for the smaller softening case. 
Our results are thus converged in time, resolution, and softening length.

Figure \ref{fig:mdot_time} plots the azimuthally and time-averaged radial profiles of the mass accretion rate in the disk, $\dot{M}(\delta r) = 2 \pi r \avg{ \Sigma \delta v_r}$, for two $\gamma_{\rm CSD}=1$ simulations. 
We compute $\dot{M}$ by stacking the 2D profiles of the radial momentum in the frame centered on the BBH center of mass, $\Sigma \delta v_r$, and take the time and azimuthal average. 
We find that outside of $\delta r > a_{\rm bin}$, the $\dot{M}$ profiles are both constant and equal to zero. 
For a nonaccreting BBH, $\dot{M}(R_{\rm H}) \approx 0$ indicates that the mass in the Hill sphere has reached a steady-state value, and together with Figure \ref{fig:adot_time} verifies that our simulations are converged in time.

\end{appendix}

\bibliography{references.bib}{}

\begin{thebibliography}{}
\expandafter\ifx\csname natexlab\endcsname\relax\def\natexlab#1{#1}\fi
\providecommand{\url}[1]{\href{#1}{#1}}
\providecommand{\dodoi}[1]{doi:~\href{http://doi.org/#1}{\nolinkurl{#1}}}
\providecommand{\doeprint}[1]{\href{http://ascl.net/#1}{\nolinkurl{http://ascl.net/#1}}}
\providecommand{\doarXiv}[1]{\href{https://arxiv.org/abs/#1}{\nolinkurl{https://arxiv.org/abs/#1}}}

\bibitem[{{Abbott} {et~al.}(2020){Abbott}, {Abbott}, {Abraham}, {Acernese},
  {Ackley}, {Adams}, {Adhikari}, {Adya}, {Affeldt}, {Agathos}, {Agatsuma},
  {Aggarwal}, {Aguiar}, {Aich}, {Aiello}, {Ain}, {Ajith}, {Akcay}, {Allen},
  {Allocca}, {Altin}, {Amato}, {Anand}, {Ananyeva}, {Anderson}, {Anderson},
  {Angelova}, {Ansoldi}, {Antier}, {Appert}, {Arai}, {Araya}, {Areeda},
  {Ar{\`e}ne}, {Arnaud}, {Aronson}, {Arun}, {Asali}, {Ascenzi}, {Ashton},
  {Aston}, {Astone}, {Aubin}, {Aufmuth}, {AultONeal}, {Austin}, {Avendano},
  {Babak}, {Bacon}, {Badaracco}, {Bader}, {Bae}, {Baer}, {Baird}, {Baldaccini},
  {Ballardin}, {Ballmer}, {Bals}, {Balsamo}, {Baltus}, {Banagiri}, {Bankar},
  {Bankar}, {Barayoga}, {Barbieri}, {Barish}, {Barker}, {Barkett}, {Barneo},
  {Barone}, {Barr}, {Barsotti}, {Barsuglia}, {Barta}, {Bartlett}, {Bartos},
  {Bassiri}, {Basti}, {Bawaj}, {Bayley}, {Bazzan}, {B{\'e}csy}, {Bejger},
  {Belahcene}, {Bell}, {Beniwal}, {Benjamin}, {Bentley}, {Bergamin}, {Berger},
  {Bergmann}, {Bernuzzi}, {Berry}, {Bersanetti}, {Bertolini}, {Betzwieser},
  {Bhandare}, {Bhandari}, {Bidler}, {Biggs}, {Bilenko}, {Billingsley},
  {Birney}, {Birnholtz}, {Biscans}, {Bischi}, {Biscoveanu}, {Bisht},
  {Bissenbayeva}, {Bitossi}, {Bizouard}, {Blackburn}, {Blackman}, {Blair},
  {Blair}, {Blair}, {Bobba}, {Bode}, {Boer}, {Boetzel}, {Bogaert}, {Bondu},
  {Bonilla}, {Bonnand}, {Booker}, {Boom}, {Bork}, {Boschi}, {Bose},
  {Bossilkov}, {Bosveld}, {Bouffanais}, {Bozzi}, {Bradaschia}, {Brady},
  {Bramley}, {Branchesi}, {Brau}, {Breschi}, {Briant}, {Briggs}, {Brighenti},
  {Brillet}, {Brinkmann}, {Brockill}, {Brooks}, {Brooks}, {Brown}, {Brunett},
  {Bruno}, {Bruntz}, {Buikema}, {Bulik}, {Bulten}, {Buonanno}, {Buscicchio},
  {Buskulic}, {Byer}, {Cabero}, {Cadonati}, {Cagnoli}, {Cahillane}, {Bustillo},
  {Callaghan}, {Callister}, {Calloni}, {Camp}, {Canepa}, {Cannon}, {Cao},
  {Cao}, {Carapella}, {Carbognani}, {Caride}, {Carney}, {Carullo}, {Diaz},
  {Casentini}, {Casta{\~n}eda}, {Caudill}, {Cavagli{\`a}}, {Cavalier},
  {Cavalieri}, {Cella}, {Cerd{\'a}-Dur{\'a}n}, {Cesarini}, {Chaibi},
  {Chakravarti}, {Chan}, {Chan}, {Chao}, {Charlton}, {Chase},
  {Chassande-Mottin}, {Chatterjee}, {Chaturvedi}, {Chatziioannou}, {Chen},
  {Chen}, {Chen}, {Cheng}, {Cheong}, {Chia}, {Chiadini}, {Chierici},
  {Chincarini}, {Chiummo}, {Cho}, {Cho}, {Cho}, {Christensen}, {Chu}, {Chua},
  {Chung}, {Chung}, {Ciani}, {Ciecielag}, {Cie{\'s}lar}, {Ciobanu}, {Ciolfi},
  {Cipriano}, {Cirone}, {Clara}, {Clark}, {Clearwater}, {Clesse}, {Cleva},
  {Coccia}, {Cohadon}, {Cohen}, {Colleoni}, {Collette}, {Collins}, {Colpi},
  {Constancio}, {Conti}, {Cooper}, {Corban}, {Corbitt}, {Cordero-Carri{\'o}n},
  {Corezzi}, {Corley}, {Cornish}, {Corre}, {Corsi}, {Cortese}, {Costa},
  {Cotesta}, {Coughlin}, {Coughlin}, {Coulon}, {Countryman}, {Couvares},
  {Covas}, {Coward}, {Cowart}, {Coyne}, {Coyne}, {Creighton}, {Creighton},
  {Cripe}, {Croquette}, {Crowder}, {Cudell}, {Cullen}, {Cumming}, {Cummings},
  {Cunningham}, {Cuoco}, {Curylo}, {Canton}, {D{\'a}lya}, {Dana},
  {Daneshgaran-Bajastani}, {D'Angelo}, {Danilishin}, {D'Antonio}, {Danzmann},
  {Darsow-Fromm}, {Dasgupta}, {Datrier}, {Dattilo}, {Dave}, {Davier}, {Davies},
  {Davis}, {Daw}, {DeBra}, {Deenadayalan}, {Degallaix}, {De Laurentis},
  {Del{\'e}glise}, {Delfavero}, {De Lillo}, {Del Pozzo}, {DeMarchi},
  {D'Emilio}, {Demos}, {Dent}, {De Pietri}, {De Rosa}, {De Rossi}, {DeSalvo},
  {de Varona}, {Dhurandhar}, {D{\'\i}az}, {Diaz-Ortiz}, {Dietrich}, {Di Fiore},
  {Di Fronzo}, {Di Giorgio}, {Di Giovanni}, {Di Giovanni}, {Di Girolamo}, {Di
  Lieto}, {Ding}, {Di Pace}, {Di Palma}, {Di Renzo}, {Divakarla}, {Dmitriev},
  {Doctor}, {Donovan}, {Dooley}, {Doravari}, {Dorrington}, {Downes}, {Drago},
  {Driggers}, {Du}, {Ducoin}, {Dupej}, {Durante}, {D'Urso}, {Dwyer}, {Easter},
  {Eddolls}, {Edelman}, {Edo}, {Edy}, {Effler}, {Ehrens}, {Eichholz},
  {Eikenberry}, {Eisenmann}, {Eisenstein}, {Ejlli}, {Errico}, {Essick},
  {Estelles}, {Estevez}, {Etienne}, {Etzel}, {Evans}, {Evans}, {Ewing},
  {Fafone}, {Fairhurst}, {Fan}, {Farinon}, {Farr}, {Farr}, {Fauchon-Jones},
  {Favata}, {Fays}, {Fazio}, {Feicht}, {Fejer}, {Feng}, {Fenyvesi}, {Ferguson},
  {Fernandez-Galiana}, {Ferrante}, {Ferreira}, {Ferreira}, {Fidecaro}, {Fiori},
  {Fiorucci}, {Fishbach}, {Fisher}, {Fittipaldi}, {Fitz-Axen}, {Fiumara},
  {Flaminio}, {Floden}, {Flynn}, {Fong}, {Font}, {Forsyth}, {Fournier},
  {Frasca}, {Frasconi}, {Frei}, {Freise}, {Frey}, {Frey}, {Fritschel},
  {Frolov}, {Fronz{\`e}}, {Fulda}, {Fyffe}, {Gabbard}, {Gadre}, {Gaebel},
  {Gair}, {Galaudage}, {Ganapathy}, {Gaonkar}, {Garc{\'\i}a-Quir{\'o}s},
  {Garufi}, {Gateley}, {Gaudio}, {Gayathri}, {Gemme}, {Genin}, {Gennai},
  {George}, {George}, {Gergely}, {Ghonge}, {Ghosh}, {Ghosh}, {Ghosh},
  {Giacomazzo}, {Giaime}, {Giardina}, {Gibson}, {Gier}, {Gill}, {Glanzer},
  {Gniesmer}, {Godwin}, {Goetz}, {Goetz}, {Gohlke}, {Goncharov},
  {Gonz{\'a}lez}, {Gopakumar}, {Gossan}, {Gosselin}, {Gouaty}, {Grace},
  {Grado}, {Granata}, {Grant}, {Gras}, {Grassia}, {Gray}, {Gray}, {Greco},
  {Green}, {Green}, {Gretarsson}, {Griggs}, {Grignani}, {Grimaldi}, {Grimm},
  {Grote}, {Grunewald}, {Gruning}, {Guidi}, {Guimaraes}, {Guix{\'e}}, {Gulati},
  {Guo}, {Gupta}, {Gupta}, {Gupta}, {Gustafson}, {Gustafson}, {Haegel},
  {Halim}, {Hall}, {Hamilton}, {Hammond}, {Haney}, {Hanke}, {Hanks}, {Hanna},
  {Hannam}, {Hannuksela}, {Hansen}, {Hanson}, {Harder}, {Hardwick}, {Haris},
  {Harms}, {Harry}, {Harry}, {Hasskew}, {Haster}, {Haughian}, {Hayes}, {Healy},
  {Heidmann}, {Heintze}, {Heinze}, {Heitmann}, {Hellman}, {Hello}, {Hemming},
  {Hendry}, {Heng}, {Hennes}, {Hennig}, {Heurs}, {Hild}, {Hinderer}, {Hoback},
  {Hochheim}, {Hofgard}, {Hofman}, {Holgado}, {Holland}, {Holt}, {Holz},
  {Hopkins}, {Horst}, {Hough}, {Howell}, {Hoy}, {Huang}, {H{\"u}bner},
  {Huerta}, {Huet}, {Hughey}, {Hui}, {Husa}, {Huttner}, {Huxford},
  {Huynh-Dinh}, {Idzkowski}, {Iess}, {Inchauspe}, {Ingram}, {Intini}, {Isac},
  {Isi}, {Iyer}, {Jacqmin}, {Jadhav}, {Jadhav}, {James}, {Jani}, {Janthalur},
  {Jaranowski}, {Jariwala}, {Jaume}, {Jenkins}, {Jiang}, {Johns},
  {Johnson-McDaniel}, {Jones}, {Jones}, {Jones}, {Jones}, {Jones}, {Jonker},
  {Ju}, {Junker}, {Kalaghatgi}, {Kalogera}, {Kamai}, {Kandhasamy}, {Kang},
  {Kanner}, {Kapadia}, {Karki}, {Kashyap}, {Kasprzack}, {Kastaun},
  {Katsanevas}, {Katsavounidis}, {Katzman}, {Kaufer}, {Kawabe},
  {K{\'e}f{\'e}lian}, {Keitel}, {Keivani}, {Kennedy}, {Key}, {Khadka},
  {Khalili}, {Khan}, {Khan}, {Khan}, {Khazanov}, {Khetan}, {Khursheed},
  {Kijbunchoo}, {Kim}, {Kim}, {Kim}, {Kim}, {Kim}, {Kim}, {Kim}, {Kimball},
  {King}, {Kinley-Hanlon}, {Kirchhoff}, {Kissel}, {Kleybolte}, {Klimenko},
  {Knowles}, {Knyazev}, {Koch}, {Koehlenbeck}, {Koekoek}, {Koley},
  {Kondrashov}, {Kontos}, {Koper}, {Korobko}, {Korth}, {Kovalam}, {Kozak},
  {Kringel}, {Krishnendu}, {Kr{\'o}lak}, {Krupinski}, {Kuehn}, {Kumar},
  {Kumar}, {Kumar}, {Kumar}, {Kumar}, {Kuo}, {Kutynia}, {Lackey}, {Laghi},
  {Lalande}, {Lam}, {Lamberts}, {Landry}, {Lane}, {Lang}, {Lange}, {Lantz},
  {Lanza}, {La Rosa}, {Lartaux-Vollard}, {Lasky}, {Laxen}, {Lazzarini},
  {Lazzaro}, {Leaci}, {Leavey}, {Lecoeuche}, {Lee}, {Lee}, {Lee}, {Lee}, {Lee},
  {Lehmann}, {Leroy}, {Letendre}, {Levin}, {Li}, {Li}, {li}, {Li}, {Li},
  {Linde}, {Linker}, {Linley}, {Littenberg}, {Liu}, {Liu},
  {Llorens-Monteagudo}, {Lo}, {Lockwood}, {London}, {Longo}, {Lorenzini},
  {Loriette}, {Lormand}, {Losurdo}, {Lough}, {Lousto}, {Lovelace}, {L{\"u}ck},
  {Lumaca}, {Lundgren}, {Ma}, {Macas}, {Macfoy}, {MacInnis}, {Macleod},
  {MacMillan}, {Macquet}, {Hernandez}, {Maga{\~n}a-Sandoval}, {Magee},
  {Majorana}, {Maksimovic}, {Malik}, {Man}, {Mandic}, {Mangano}, {Mansell},
  {Manske}, {Mantovani}, {Mapelli}, {Marchesoni}, {Marion}, {M{\'a}rka},
  {M{\'a}rka}, {Markakis}, {Markosyan}, {Markowitz}, {Maros}, {Marquina},
  {Marsat}, {Martelli}, {Martin}, {Martin}, {Martinez}, {Martynov},
  {Masalehdan}, {Mason}, {Massera}, {Masserot}, {Massinger}, {Masso-Reid},
  {Mastrogiovanni}, {Matas}, {Matichard}, {Mavalvala}, {Maynard}, {McCann},
  {McCarthy}, {McClelland}, {McCormick}, {McCuller}, {McGuire}, {McIsaac},
  {McIver}, {McManus}, {McRae}, {McWilliams}, {Meacher}, {Meadors}, {Mehmet},
  {Mehta}, {Villa}, {Melatos}, {Mendell}, {Mercer}, {Mereni}, {Merfeld},
  {Merilh}, {Merritt}, {Merzougui}, {Meshkov}, {Messenger}, {Messick},
  {Metzdorff}, {Meyers}, {Meylahn}, {Mhaske}, {Miani}, {Miao}, {Michaloliakos},
  {Michel}, {Middleton}, {Milano}, {Miller}, {Millhouse}, {Mills}, {Milotti},
  {Milovich-Goff}, {Minazzoli}, {Minenkov}, {Mishkin}, {Mishra}, {Mistry},
  {Mitra}, {Mitrofanov}, {Mitselmakher}, {Mittleman}, {Mo}, {Mogushi},
  {Mohapatra}, {Mohite}, {Molina-Ruiz}, {Mondin}, {Montani}, {Moore}, {Moraru},
  {Morawski}, {Moreno}, {Morisaki}, {Mours}, {Mow-Lowry}, {Mozzon},
  {Muciaccia}, {Mukherjee}, {Mukherjee}, {Mukherjee}, {Mukherjee}, {Mukund},
  {Mullavey}, {Munch}, {Mu{\~n}iz}, {Murray}, {Nagar}, {Nardecchia},
  {Naticchioni}, {Nayak}, {Neil}, {Neilson}, {Nelemans}, {Nelson}, {Nery},
  {Neunzert}, {Ng}, {Ng}, {Nguyen}, {Nguyen}, {Nichols}, {Nichols}, {Nissanke},
  {Nocera}, {Noh}, {North}, {Nothard}, {Nuttall}, {Oberling}, {O'Brien},
  {Oganesyan}, {Ogin}, {Oh}, {Oh}, {Ohme}, {Ohta}, {Okada}, {Oliver},
  {Olivetto}, {Oppermann}, {Oram}, {O'Reilly}, {Ormiston}, {Ortega},
  {O'Shaughnessy}, {Ossokine}, {Osthelder}, {Ottaway}, {Overmier}, {Owen},
  {Pace}, {Pagano}, {Page}, {Pagliaroli}, {Pai}, {Pai}, {Palamos}, {Palashov},
  {Palomba}, {Pan}, {Panda}, {Pang}, {Pankow}, {Pannarale}, {Pant}, {Paoletti},
  {Paoli}, {Parida}, {Parker}, {Pascucci}, {Pasqualetti}, {Passaquieti},
  {Passuello}, {Patricelli}, {Payne}, {Pearlstone}, {Pechsiri}, {Pedersen},
  {Pedraza}, {Pele}, {Penn}, {Perego}, {Perez}, {P{\'e}rigois}, {Perreca},
  {Perri{\`e}s}, {Petermann}, {Pfeiffer}, {Phelps}, {Phukon}, {Piccinni},
  {Pichot}, {Piendibene}, {Piergiovanni}, {Pierro}, {Pillant}, {Pinard},
  {Pinto}, {Piotrzkowski}, {Pirello}, {Pitkin}, {Plastino}, {Poggiani}, {Pong},
  {Ponrathnam}, {Popolizio}, {Porter}, {Powell}, {Prajapati}, {Prasai},
  {Prasanna}, {Pratten}, {Prestegard}, {Principe}, {Prodi}, {Prokhorov},
  {Punturo}, {Puppo}, {P{\"u}rrer}, {Qi}, {Quetschke}, {Quinonez}, {Raab},
  {Raaijmakers}, {Radkins}, {Radulesco}, {Raffai}, {Rafferty}, {Raja}, {Rajan},
  {Rajbhandari}, {Rakhmanov}, {Ramirez}, {Ramos-Buades}, {Rana}, {Rao},
  {Rapagnani}, {Raymond}, {Razzano}, {Read}, {Regimbau}, {Rei}, {Reid},
  {Reitze}, {Rettegno}, {Ricci}, {Richardson}, {Richardson}, {Ricker},
  {Riemenschneider}, {Riles}, {Rizzo}, {Robertson}, {Robinet}, {Rocchi},
  {Rodriguez-Soto}, {Rolland}, {Rollins}, {Roma}, {Romanelli}, {Romano},
  {Romel}, {Romero-Shaw}, {Romie}, {Rose}, {Rose}, {Rose}, {Rosi{\'n}ska},
  {Rosofsky}, {Ross}, {Rowan}, {Rowlinson}, {Roy}, {Roy}, {Roy}, {Ruggi},
  {Rutins}, {Ryan}, {Sachdev}, {Sadecki}, {Sakellariadou}, {Salafia},
  {Salconi}, {Saleem}, {Samajdar}, {Sanchez}, {Sanchez}, {Sanchis-Gual},
  {Sanders}, {Santiago}, {Santos}, {Sarin}, {Sassolas}, {Sathyaprakash},
  {Sauter}, {Savage}, {Savant}, {Sawant}, {Sayah}, {Schaetzl}, {Schale},
  {Scheel}, {Scheuer}, {Schmidt}, {Schnabel}, {Schofield}, {Sch{\"o}nbeck},
  {Schreiber}, {Schulte}, {Schutz}, {Schwarm}, {Schwartz}, {Scott}, {Scott},
  {Seidel}, {Sellers}, {Sengupta}, {Sennett}, {Sentenac}, {Sequino}, {Sergeev},
  {Setyawati}, {Shaddock}, {Shaffer}, {Shahriar}, {Sharifi}, {Sharma},
  {Sharma}, {Shawhan}, {Shen}, {Shikauchi}, {Shink}, {Shoemaker}, {Shoemaker},
  {Shukla}, {ShyamSundar}, {Siellez}, {Sieniawska}, {Sigg}, {Singer}, {Singh},
  {Singh}, {Singha}, {Singhal}, {Sintes}, {Sipala}, {Skliris}, {Slagmolen},
  {Slaven-Blair}, {Smetana}, {Smith}, {Smith}, {Somala}, {Son}, {Soni},
  {Sorazu}, {Sordini}, {Sorrentino}, {Souradeep}, {Sowell}, {Spencer}, {Spera},
  {Srivastava}, {Srivastava}, {Staats}, {Stachie}, {Standke}, {Steer},
  {Steinke}, {Steinlechner}, {Steinlechner}, {Steinmeyer}, {Stevenson},
  {Stocks}, {Stops}, {Stover}, {Strain}, {Stratta}, {Strunk}, {Sturani},
  {Stuver}, {Sudhagar}, {Sudhir}, {Summerscales}, {Sun}, {Sunil}, {Sur},
  {Suresh}, {Sutton}, {Swinkels}, {Szczepa{\'n}czyk}, {Tacca}, {Tait},
  {Talbot}, {Tanasijczuk}, {Tanner}, {Tao}, {T{\'a}pai}, {Tapia}, {San Martin},
  {Tasson}, {Taylor}, {Tenorio}, {Terkowski}, {Thirugnanasambandam}, {Thomas},
  {Thomas}, {Thompson}, {Thondapu}, {Thorne}, {Thrane}, {Tinsman}, {Saravanan},
  {Tiwari}, {Tiwari}, {Tiwari}, {Toland}, {Tonelli}, {Tornasi},
  {Torres-Forn{\'e}}, {Torrie}, {Tosta e Melo}, {T{\"o}yr{\"a}}, {Trail},
  {Travasso}, {Traylor}, {Tringali}, {Tripathee}, {Trovato}, {Trudeau},
  {Tsang}, {Tse}, {Tso}, {Tsukada}, {Tsuna}, {Tsutsui}, {Turconi}, {Ubhi},
  {Udall}, {Ueno}, {Ugolini}, {Unnikrishnan}, {Urban}, {Usman}, {Utina},
  {Vahlbruch}, {Vajente}, {Valdes}, {Valentini}, {van Bakel}, {van Beuzekom},
  {van den Brand}, {Van Den Broeck}, {Vander-Hyde}, {van der Schaaf}, {Van
  Heijningen}, {van Veggel}, {Vardaro}, {Varma}, {Vass}, {Vas{\'u}th},
  {Vecchio}, {Vedovato}, {Veitch}, {Veitch}, {Venkateswara}, {Venugopalan},
  {Verkindt}, {Veske}, {Vetrano}, {Vicer{\'e}}, {Viets}, {Vinciguerra}, {Vine},
  {Vinet}, {Vitale}, {Vivanco}, {Vo}, {Vocca}, {Vorvick}, {Vyatchanin}, {Wade},
  {Wade}, {Wade}, {Walet}, {Walker}, {Wallace}, {Wallace}, {Walsh}, {Wang},
  {Wang}, {Wang}, {Ward}, {Warden}, {Warner}, {Was}, {Watchi}, {Weaver}, {Wei},
  {Weinert}, {Weinstein}, {Weiss}, {Wellmann}, {Wen}, {We{\ss}els},
  {Westhouse}, {Wette}, {Whelan}, {Whiting}, {Whittle}, {Wilken}, {Williams},
  {Willis}, {Willke}, {Winkler}, {Wipf}, {Wittel}, {Woan}, {Woehler},
  {Wofford}, {Wong}, {Wright}, {Wu}, {Wysocki}, {Xiao}, {Yamamoto}, {Yang},
  {Yang}, {Yang}, {Yap}, {Yazback}, {Yeeles}, {Yu}, {Yu}, {Yuen},
  {Zadro{\.z}ny}, {Zadro{\.z}ny}, {Zanolin}, {Zelenova}, {Zendri}, {Zevin},
  {Zhang}, {Zhang}, {Zhang}, {Zhao}, {Zhao}, {Zhou}, {Zhou}, {Zhu},
  {Zimmerman}, {Zlochower}, {Zucker}, {Zweizig}, {LIGO Scientific
  Collaboration}, \& {Virgo Collaboration}}]{Abbott2020b}
{Abbott}, R., {Abbott}, T.~D., {Abraham}, S., {et~al.} 2020,
  \href{http://dx.doi.org/10.3847/2041-8213/aba493}{\apjl},
  \href{https://ui.adsabs.harvard.edu/abs/2020ApJ...900L..13A}{900, L13}

\bibitem[{{Abramowicz} {et~al.}(1980){Abramowicz}, {Calvani}, \&
  {Nobili}}]{Abramowicz1980}
{Abramowicz}, M.~A., {Calvani}, M., \& {Nobili}, L. 1980,
  \href{http://dx.doi.org/10.1086/158512}{\apj},
  \href{https://ui.adsabs.harvard.edu/abs/1980ApJ...242..772A}{242, 772}

\bibitem[{{Artymowicz} \& {Lubow}(1994)}]{Artymowicz.1994}
{Artymowicz}, P., \& {Lubow}, S.~H. 1994,
  \href{http://dx.doi.org/10.1086/173679}{\apj},
  \href{https://ui.adsabs.harvard.edu/abs/1994ApJ...421..651A}{421, 651}

\bibitem[{{Bartos} {et~al.}(2017){Bartos}, {Kocsis}, {Haiman}, \&
  {M{\'a}rka}}]{Bartos2017}
{Bartos}, I., {Kocsis}, B., {Haiman}, Z., \& {M{\'a}rka}, S. 2017,
  \href{http://dx.doi.org/10.3847/1538-4357/835/2/165}{\apj},
  \href{https://ui.adsabs.harvard.edu/abs/2017ApJ...835..165B}{835, 165}

\bibitem[{{Baruteau} {et~al.}(2011){Baruteau}, {Cuadra}, \&
  {Lin}}]{Baruteau2011}
{Baruteau}, C., {Cuadra}, J., \& {Lin}, D.~N.~C. 2011,
  \href{http://dx.doi.org/10.1088/0004-637X/726/1/28}{\apj},
  \href{https://ui.adsabs.harvard.edu/abs/2011ApJ...726...28B}{726, 28}

\bibitem[{{Belczynski} {et~al.}(2016){Belczynski}, {Holz}, {Bulik}, \&
  {O'Shaughnessy}}]{Belczynski2016}
{Belczynski}, K., {Holz}, D.~E., {Bulik}, T., \& {O'Shaughnessy}, R. 2016,
  \href{http://dx.doi.org/10.1038/nature18322}{\nat},
  \href{https://ui.adsabs.harvard.edu/abs/2016Natur.534..512B}{534, 512}

\bibitem[{{Benacquista} \& {Downing}(2013)}]{Benacquista2013}
{Benacquista}, M.~J., \& {Downing}, J. M.~B. 2013,
  \href{http://dx.doi.org/10.12942/lrr-2013-4}{Living Reviews in Relativity},
  \href{https://ui.adsabs.harvard.edu/abs/2013LRR....16....4B}{16, 4}

\bibitem[{{de Val-Borro} {et~al.}(2006){de Val-Borro}, {Edgar}, {Artymowicz},
  {Ciecielag}, {Cresswell}, {D'Angelo}, {Delgado-Donate}, {Dirksen}, {Fromang},
  {Gawryszczak}, {Klahr}, {Kley}, {Lyra}, {Masset}, {Mellema}, {Nelson},
  {Paardekooper}, {Peplinski}, {Pierens}, {Plewa}, {Rice}, {Sch{\"a}fer}, \&
  {Speith}}]{deValborro2006}
{de Val-Borro}, M., {Edgar}, R.~G., {Artymowicz}, P., {et~al.} 2006,
  \href{http://dx.doi.org/10.1111/j.1365-2966.2006.10488.x}{\mnras},
  \href{https://ui.adsabs.harvard.edu/abs/2006MNRAS.370..529D}{370, 529}

\bibitem[{{del Valle} \& {Volonteri}(2018)}]{delValle2018}
{del Valle}, L., \& {Volonteri}, M. 2018,
  \href{http://dx.doi.org/10.1093/mnras/sty1815}{\mnras},
  \href{https://ui.adsabs.harvard.edu/abs/2018MNRAS.480..439D}{480, 439}

\bibitem[{{Dempsey} {et~al.}(2020){Dempsey}, {Lee}, \&
  {Lithwick}}]{Dempsey2020}
{Dempsey}, A.~M., {Lee}, W.-K., \& {Lithwick}, Y. 2020,
  \href{http://dx.doi.org/10.3847/1538-4357/ab723c}{\apj},
  \href{https://ui.adsabs.harvard.edu/abs/2020ApJ...891..108D}{891, 108}

\bibitem[{{Dempsey} {et~al.}(2021){Dempsey}, {Mu{\~n}oz}, \&
  {Lithwick}}]{Dempsey2021}
{Dempsey}, A.~M., {Mu{\~n}oz}, D.~J., \& {Lithwick}, Y. 2021,
  \href{http://dx.doi.org/10.3847/2041-8213/ac22af}{\apjl},
  \href{https://ui.adsabs.harvard.edu/abs/2021ApJ...918L..36D}{918, L36}

\bibitem[{{Dittmann} \& {Ryan}(2021)}]{Dittmann2021}
{Dittmann}, A.~J., \& {Ryan}, G. 2021,
  \href{http://dx.doi.org/10.3847/1538-4357/ac1bbd}{\apj},
  \href{https://ui.adsabs.harvard.edu/abs/2021ApJ...921...71D}{921, 71}

\bibitem[{{Dittmann} \& {Ryan}(2022)}]{Dittmann2022}
---. 2022, \href{https://ui.adsabs.harvard.edu/abs/2022arXiv220107816D}{arXiv
  e-prints, arXiv:2201.07816}

\bibitem[{{D'Orazio} \& {Duffell}(2021)}]{DOrazio2021}
{D'Orazio}, D.~J., \& {Duffell}, P.~C. 2021,
  \href{http://dx.doi.org/10.3847/2041-8213/ac0621}{\apjl},
  \href{https://ui.adsabs.harvard.edu/abs/2021ApJ...914L..21D}{914, L21}

\bibitem[{{Duffell} {et~al.}(2020){Duffell}, {D'Orazio}, {Derdzinski},
  {Haiman}, {MacFadyen}, {Rosen}, \& {Zrake}}]{Duffell2020}
{Duffell}, P.~C., {D'Orazio}, D., {Derdzinski}, A., {et~al.} 2020,
  \href{http://dx.doi.org/10.3847/1538-4357/abab95}{\apj},
  \href{https://ui.adsabs.harvard.edu/abs/2020ApJ...901...25D}{901, 25}

\bibitem[{{Fern{\'a}ndez} \& {Kobayashi}(2019)}]{Fernandez2019}
{Fern{\'a}ndez}, J.~J., \& {Kobayashi}, S. 2019,
  \href{http://dx.doi.org/10.1093/mnras/stz1353}{\mnras},
  \href{https://ui.adsabs.harvard.edu/abs/2019MNRAS.487.1200F}{487, 1200}

\bibitem[{{Fragione} \& {Kocsis}(2019)}]{Fragione2019}
{Fragione}, G., \& {Kocsis}, B. 2019,
  \href{http://dx.doi.org/10.1093/mnras/stz1175}{\mnras},
  \href{https://ui.adsabs.harvard.edu/abs/2019MNRAS.486.4781F}{486, 4781}

\bibitem[{{Fragione} {et~al.}(2022){Fragione}, {Kocsis}, {Rasio}, \&
  {Silk}}]{Fragione2022}
{Fragione}, G., {Kocsis}, B., {Rasio}, F.~A., \& {Silk}, J. 2022,
  \href{http://dx.doi.org/10.3847/1538-4357/ac5026}{\apj},
  \href{https://ui.adsabs.harvard.edu/abs/2022ApJ...927..231F}{927, 231}

\bibitem[{{Frank} {et~al.}(1992){Frank}, {King}, \& {Raine}}]{Frank1992}
{Frank}, J., {King}, A., \& {Raine}, D. 1992, {Accretion power in
  astrophysics.}, Vol.~21

\bibitem[{{Graham} {et~al.}(2020){Graham}, {Ford}, {McKernan}, {Ross}, {Stern},
  {Burdge}, {Coughlin}, {Djorgovski}, {Drake}, {Duev}, {Kasliwal}, {Mahabal},
  {van Velzen}, {Belecki}, {Bellm}, {Burruss}, {Cenko}, {Cunningham}, {Helou},
  {Kulkarni}, {Masci}, {Prince}, {Reiley}, {Rodriguez}, {Rusholme}, {Smith}, \&
  {Soumagnac}}]{Graham2020}
{Graham}, M.~J., {Ford}, K.~E.~S., {McKernan}, B., {et~al.} 2020,
  \href{http://dx.doi.org/10.1103/PhysRevLett.124.251102}{\prl},
  \href{https://ui.adsabs.harvard.edu/abs/2020PhRvL.124y1102G}{124, 251102}

\bibitem[{{Gr{\"o}bner} {et~al.}(2020){Gr{\"o}bner}, {Ishibashi}, {Tiwari},
  {Haney}, \& {Jetzer}}]{Grobner2020}
{Gr{\"o}bner}, M., {Ishibashi}, W., {Tiwari}, S., {Haney}, M., \& {Jetzer}, P.
  2020, \href{http://dx.doi.org/10.1051/0004-6361/202037681}{\aap},
  \href{https://ui.adsabs.harvard.edu/abs/2020A&A...638A.119G}{638, A119}

\bibitem[{{Hankla} {et~al.}(2020){Hankla}, {Jiang}, \& {Armitage}}]{Hankla2020}
{Hankla}, A.~M., {Jiang}, Y.-F., \& {Armitage}, P.~J. 2020,
  \href{http://dx.doi.org/10.3847/1538-4357/abb4df}{\apj},
  \href{https://ui.adsabs.harvard.edu/abs/2020ApJ...902...50H}{902, 50}

\bibitem[{{Heath} \& {Nixon}(2020)}]{Heath2020}
{Heath}, R.~M., \& {Nixon}, C.~J. 2020,
  \href{http://dx.doi.org/10.1051/0004-6361/202038548}{\aap},
  \href{https://ui.adsabs.harvard.edu/abs/2020A&A...641A..64H}{641, A64}

\bibitem[{{Hunter}(2007)}]{Hunter2007}
{Hunter}, J.~D. 2007, \href{http://dx.doi.org/10.1109/MCSE.2007.55}{Computing
  in Science and Engineering},
  \href{https://ui.adsabs.harvard.edu/abs/2007CSE.....9...90H}{9, 90}

\bibitem[{{Jiang} {et~al.}(2014){Jiang}, {Stone}, \& {Davis}}]{Jiang2014}
{Jiang}, Y.-F., {Stone}, J.~M., \& {Davis}, S.~W. 2014,
  \href{http://dx.doi.org/10.1088/0004-637X/796/2/106}{\apj},
  \href{https://ui.adsabs.harvard.edu/abs/2014ApJ...796..106J}{796, 106}

\bibitem[{{Kaaz} {et~al.}(2021){Kaaz}, {Schr{\o}der}, {Andrews}, {Antoni}, \&
  {Ramirez-Ruiz}}]{Kaaz2021}
{Kaaz}, N., {Schr{\o}der}, S.~L., {Andrews}, J.~J., {Antoni}, A., \&
  {Ramirez-Ruiz}, E. 2021,
  \href{https://ui.adsabs.harvard.edu/abs/2021arXiv210312088K}{arXiv e-prints,
  arXiv:2103.12088}

\bibitem[{{Kimura} {et~al.}(2021){Kimura}, {Murase}, \& {Bartos}}]{Kimura2021}
{Kimura}, S.~S., {Murase}, K., \& {Bartos}, I. 2021,
  \href{http://dx.doi.org/10.3847/1538-4357/ac0535}{\apj},
  \href{https://ui.adsabs.harvard.edu/abs/2021ApJ...916..111K}{916, 111}

\bibitem[{{Kley} {et~al.}(2009){Kley}, {Bitsch}, \& {Klahr}}]{Kley2009}
{Kley}, W., {Bitsch}, B., \& {Klahr}, H. 2009,
  \href{http://dx.doi.org/10.1051/0004-6361/200912072}{\aap},
  \href{https://ui.adsabs.harvard.edu/abs/2009A&A...506..971K}{506, 971}

\bibitem[{{Leigh} {et~al.}(2018){Leigh}, {Geller}, {McKernan}, {Ford}, {Mac
  Low}, {Bellovary}, {Haiman}, {Lyra}, {Samsing}, {O'Dowd}, {Kocsis}, \&
  {Endlich}}]{Leigh2018}
{Leigh}, N.~W.~C., {Geller}, A.~M., {McKernan}, B., {et~al.} 2018,
  \href{http://dx.doi.org/10.1093/mnras/stx3134}{\mnras},
  \href{https://ui.adsabs.harvard.edu/abs/2018MNRAS.474.5672L}{474, 5672}

\bibitem[{{Li} {et~al.}(2005){Li}, {Li}, {Koller}, {Wendroff}, {Liska},
  {Orban}, {Liang}, \& {Lin}}]{Li2005}
{Li}, H., {Li}, S., {Koller}, J., {et~al.} 2005,
  \href{http://dx.doi.org/10.1086/429367}{\apj},
  \href{https://ui.adsabs.harvard.edu/abs/2005ApJ...624.1003L}{624, 1003}

\bibitem[{{Li} {et~al.}(2009){Li}, {Lubow}, {Li}, \& {Lin}}]{Li2009}
{Li}, H., {Lubow}, S.~H., {Li}, S., \& {Lin}, D.~N.~C. 2009,
  \href{http://dx.doi.org/10.1088/0004-637X/690/1/L52}{\apjl},
  \href{https://ui.adsabs.harvard.edu/abs/2009ApJ...690L..52L}{690, L52}

\bibitem[{{Li} \& {Lai}(2022)}]{LiLai2022}
{Li}, R., \& {Lai}, D. 2022,
  \href{https://ui.adsabs.harvard.edu/abs/2022arXiv220207633L}{arXiv e-prints,
  arXiv:2202.07633}

\bibitem[{{Li} {et~al.}(2021){Li}, {Dempsey}, {Li}, {Li}, \& {Li}}]{Li2021}
{Li}, Y.-P., {Dempsey}, A.~M., {Li}, S., {Li}, H., \& {Li}, J. 2021,
  \href{http://dx.doi.org/10.3847/1538-4357/abed48}{\apj},
  \href{https://ui.adsabs.harvard.edu/abs/2021ApJ...911..124L}{911, 124}

\bibitem[{{Li} {et~al.}(2019){Li}, {Yuan}, \& {Dai}}]{Li2019}
{Li}, Y.-P., {Yuan}, F., \& {Dai}, X. 2019,
  \href{http://dx.doi.org/10.1093/mnras/sty3245}{\mnras},
  \href{https://ui.adsabs.harvard.edu/abs/2019MNRAS.483.2275L}{483, 2275}

\bibitem[{{Lin} \& {Papaloizou}(1993)}]{Lin1993}
{Lin}, D.~N.~C., \& {Papaloizou}, J.~C.~B. 1993, in Protostars and Planets III,
  ed. E.~H. {Levy} \& J.~I. {Lunine}, 749

\bibitem[{{Liu} \& {Lai}(2021)}]{Liu2021}
{Liu}, B., \& {Lai}, D. 2021,
  \href{http://dx.doi.org/10.1093/mnras/stab178}{\mnras},
  \href{https://ui.adsabs.harvard.edu/abs/2021MNRAS.502.2049L}{502, 2049}

\bibitem[{{Mandel} \& {de Mink}(2016)}]{Mandel2016}
{Mandel}, I., \& {de Mink}, S.~E. 2016,
  \href{http://dx.doi.org/10.1093/mnras/stw379}{\mnras},
  \href{https://ui.adsabs.harvard.edu/abs/2016MNRAS.458.2634M}{458, 2634}

\bibitem[{{Masset}(2017)}]{Masset2017}
{Masset}, F.~S. 2017, \href{http://dx.doi.org/10.1093/mnras/stx2271}{\mnras},
  \href{https://ui.adsabs.harvard.edu/abs/2017MNRAS.472.4204M}{472, 4204}

\bibitem[{{McKernan} {et~al.}(2021){McKernan}, {Ford}, {Callister}, {Farr},
  {O'Shaughnessy}, {Smith}, {Thrane}, \& {Vajpeyi}}]{McKernan2021}
{McKernan}, B., {Ford}, K.~E.~S., {Callister}, T., {et~al.} 2021,
  \href{https://ui.adsabs.harvard.edu/abs/2021arXiv210707551M}{arXiv e-prints,
  arXiv:2107.07551}

\bibitem[{{McKernan} {et~al.}(2012){McKernan}, {Ford}, {Lyra}, \&
  {Perets}}]{McKernan2012}
{McKernan}, B., {Ford}, K.~E.~S., {Lyra}, W., \& {Perets}, H.~B. 2012,
  \href{http://dx.doi.org/10.1111/j.1365-2966.2012.21486.x}{\mnras},
  \href{https://ui.adsabs.harvard.edu/abs/2012MNRAS.425..460M}{425, 460}

\bibitem[{{Moody} {et~al.}(2019){Moody}, {Shi}, \& {Stone}}]{Moody2019}
{Moody}, M. S.~L., {Shi}, J.-M., \& {Stone}, J.~M. 2019,
  \href{http://dx.doi.org/10.3847/1538-4357/ab09ee}{\apj},
  \href{https://ui.adsabs.harvard.edu/abs/2019ApJ...875...66M}{875, 66}

\bibitem[{{Mu{\~n}oz} {et~al.}(2020){Mu{\~n}oz}, {Lai}, {Kratter}, \& {Mirand
  a}}]{Munoz2020}
{Mu{\~n}oz}, D.~J., {Lai}, D., {Kratter}, K., \& {Mirand a}, R. 2020,
  \href{http://dx.doi.org/10.3847/1538-4357/ab5d33}{\apj},
  \href{https://ui.adsabs.harvard.edu/abs/2020ApJ...889..114M}{889, 114}

\bibitem[{{Mu{\~n}oz} {et~al.}(2019){Mu{\~n}oz}, {Miranda}, \&
  {Lai}}]{Munoz2019}
{Mu{\~n}oz}, D.~J., {Miranda}, R., \& {Lai}, D. 2019,
  \href{http://dx.doi.org/10.3847/1538-4357/aaf867}{\apj},
  \href{https://ui.adsabs.harvard.edu/abs/2019ApJ...871...84M}{871, 84}

\bibitem[{{Netzer}(2015)}]{Netzer2015}
{Netzer}, H. 2015,
  \href{http://dx.doi.org/10.1146/annurev-astro-082214-122302}{\araa},
  \href{https://ui.adsabs.harvard.edu/abs/2015ARA&A..53..365N}{53, 365}

\bibitem[{{O'Leary} {et~al.}(2009){O'Leary}, {Kocsis}, \& {Loeb}}]{OLeary2009}
{O'Leary}, R.~M., {Kocsis}, B., \& {Loeb}, A. 2009,
  \href{http://dx.doi.org/10.1111/j.1365-2966.2009.14653.x}{\mnras},
  \href{https://ui.adsabs.harvard.edu/abs/2009MNRAS.395.2127O}{395, 2127}

\bibitem[{{Proga} \& {Kallman}(2002)}]{Proga2002}
{Proga}, D., \& {Kallman}, T.~R. 2002,
  \href{http://dx.doi.org/10.1086/324534}{\apj},
  \href{https://ui.adsabs.harvard.edu/abs/2002ApJ...565..455P}{565, 455}

\bibitem[{{Rodriguez} {et~al.}(2016){Rodriguez}, {Chatterjee}, \&
  {Rasio}}]{Rodriguez2016}
{Rodriguez}, C.~L., {Chatterjee}, S., \& {Rasio}, F.~A. 2016,
  \href{http://dx.doi.org/10.1103/PhysRevD.93.084029}{\prd},
  \href{https://ui.adsabs.harvard.edu/abs/2016PhRvD..93h4029R}{93, 084029}

\bibitem[{{Samsing} {et~al.}(2020){Samsing}, {Bartos}, {D'Orazio}, {Haiman},
  {Kocsis}, {Leigh}, {Liu}, {Pessah}, \& {Tagawa}}]{Samsing2020}
{Samsing}, J., {Bartos}, I., {D'Orazio}, D.~J., {et~al.} 2020,
  \href{https://ui.adsabs.harvard.edu/abs/2020arXiv201009765S}{arXiv e-prints,
  arXiv:2010.09765}

\bibitem[{{Secunda} {et~al.}(2019){Secunda}, {Bellovary}, {Mac Low}, {Ford},
  {McKernan}, {Leigh}, {Lyra}, \& {S{\'a}ndor}}]{Secunda2019}
{Secunda}, A., {Bellovary}, J., {Mac Low}, M.-M., {et~al.} 2019,
  \href{http://dx.doi.org/10.3847/1538-4357/ab20ca}{\apj},
  \href{https://ui.adsabs.harvard.edu/abs/2019ApJ...878...85S}{878, 85}

\bibitem[{{Shakura} \& {Sunyaev}(1973)}]{Shakura1973}
{Shakura}, N.~I., \& {Sunyaev}, R.~A. 1973, \aap,
  \href{https://ui.adsabs.harvard.edu/abs/1973A&A....24..337S}{500, 33}

\bibitem[{{Shi} {et~al.}(2021){Shi}, {Li}, {Yuan}, \& {Zhu}}]{Shi2021}
{Shi}, F., {Li}, Z., {Yuan}, F., \& {Zhu}, B. 2021,
  \href{http://dx.doi.org/10.1038/s41550-021-01394-0}{Nature Astronomy},
  \href{https://ui.adsabs.harvard.edu/abs/2021NatAs...5..928S}{5, 928}

\bibitem[{{Souza Lima} {et~al.}(2017){Souza Lima}, {Mayer}, {Capelo}, \&
  {Bellovary}}]{SouzaLima2017}
{Souza Lima}, R., {Mayer}, L., {Capelo}, P.~R., \& {Bellovary}, J.~M. 2017,
  \href{http://dx.doi.org/10.3847/1538-4357/aa5d19}{\apj},
  \href{https://ui.adsabs.harvard.edu/abs/2017ApJ...838...13S}{838, 13}

\bibitem[{{Stone} {et~al.}(2017){Stone}, {Metzger}, \& {Haiman}}]{Stone2017}
{Stone}, N.~C., {Metzger}, B.~D., \& {Haiman}, Z. 2017,
  \href{http://dx.doi.org/10.1093/mnras/stw2260}{\mnras},
  \href{https://ui.adsabs.harvard.edu/abs/2017MNRAS.464..946S}{464, 946}

\bibitem[{{Sun} {et~al.}(2019){Sun}, {Xue}, {Trump}, \& {Gu}}]{Sun2019}
{Sun}, M., {Xue}, Y., {Trump}, J.~R., \& {Gu}, W.-M. 2019,
  \href{http://dx.doi.org/10.1093/mnras/sty2885}{\mnras},
  \href{https://ui.adsabs.harvard.edu/abs/2019MNRAS.482.2788S}{482, 2788}

\bibitem[{{Szul{\'a}gyi} {et~al.}(2016){Szul{\'a}gyi}, {Masset}, {Lega},
  {Crida}, {Morbidelli}, \& {Guillot}}]{Szulagyi2016}
{Szul{\'a}gyi}, J., {Masset}, F., {Lega}, E., {et~al.} 2016,
  \href{http://dx.doi.org/10.1093/mnras/stw1160}{\mnras},
  \href{https://ui.adsabs.harvard.edu/abs/2016MNRAS.460.2853S}{460, 2853}

\bibitem[{{Tagawa} {et~al.}(2020){Tagawa}, {Haiman}, \& {Kocsis}}]{Tagawa2020}
{Tagawa}, H., {Haiman}, Z., \& {Kocsis}, B. 2020,
  \href{http://dx.doi.org/10.3847/1538-4357/ab9b8c}{\apj},
  \href{https://ui.adsabs.harvard.edu/abs/2020ApJ...898...25T}{898, 25}

\bibitem[{{Tang} {et~al.}(2017){Tang}, {MacFadyen}, \& {Haiman}}]{Tang2017}
{Tang}, Y., {MacFadyen}, A., \& {Haiman}, Z. 2017,
  \href{http://dx.doi.org/10.1093/mnras/stx1130}{\mnras},
  \href{https://ui.adsabs.harvard.edu/abs/2017MNRAS.469.4258T}{469, 4258}

\bibitem[{{Tiede} {et~al.}(2020){Tiede}, {Zrake}, {MacFadyen}, \&
  {Haiman}}]{Tiede2020}
{Tiede}, C., {Zrake}, J., {MacFadyen}, A., \& {Haiman}, Z. 2020,
  \href{http://dx.doi.org/10.3847/1538-4357/aba432}{\apj},
  \href{https://ui.adsabs.harvard.edu/abs/2020ApJ...900...43T}{900, 43}

\bibitem[{{Tombesi} {et~al.}(2015){Tombesi}, {Mel{\'e}ndez}, {Veilleux},
  {Reeves}, {Gonz{\'a}lez-Alfonso}, \& {Reynolds}}]{Tombesi2015}
{Tombesi}, F., {Mel{\'e}ndez}, M., {Veilleux}, S., {et~al.} 2015,
  \href{http://dx.doi.org/10.1038/nature14261}{\nat},
  \href{https://ui.adsabs.harvard.edu/abs/2015Natur.519..436T}{519, 436}

\bibitem[{{van der Walt} {et~al.}(2011){van der Walt}, {Colbert}, \&
  {Varoquaux}}]{vanderWalt2011}
{van der Walt}, S., {Colbert}, S.~C., \& {Varoquaux}, G. 2011,
  \href{http://dx.doi.org/10.1109/MCSE.2011.37}{Computing in Science and
  Engineering},
  \href{https://ui.adsabs.harvard.edu/abs/2011CSE....13b..22V}{13, 22}

\bibitem[{{Virtanen} {et~al.}(2020){Virtanen}, {Gommers}, {Oliphant},
  {Haberland}, {Reddy}, {Cournapeau}, {Burovski}, {Peterson}, {Weckesser},
  {Bright}, {van der Walt}, {Brett}, {Wilson}, {Millman}, {Mayorov}, {Nelson},
  {Jones}, {Kern}, {Larson}, {Carey}, {Polat}, {Feng}, {Moore}, {VanderPlas},
  {Laxalde}, {Perktold}, {Cimrman}, {Henriksen}, {Quintero}, {Harris},
  {Archibald}, {Ribeiro}, {Pedregosa}, {van Mulbregt}, \& {SciPy 1. 0
  Contributors}}]{Virtanen2020}
{Virtanen}, P., {Gommers}, R., {Oliphant}, T.~E., {et~al.} 2020,
  \href{http://dx.doi.org/10.1038/s41592-019-0686-2}{Nature Methods},
  \href{https://ui.adsabs.harvard.edu/abs/2020NatMe..17..261V}{17, 261}

\bibitem[{{Wang} {et~al.}(2021){Wang}, {Liu}, {Ho}, {Li}, \& {Du}}]{Wang2021}
{Wang}, J.-M., {Liu}, J.-R., {Ho}, L.~C., {Li}, Y.-R., \& {Du}, P. 2021,
  \href{http://dx.doi.org/10.3847/2041-8213/ac0b46}{\apjl},
  \href{https://ui.adsabs.harvard.edu/abs/2021ApJ...916L..17W}{916, L17}

\bibitem[{{Yang} {et~al.}(2019){Yang}, {Bartos}, {Gayathri}, {Ford}, {Haiman},
  {Klimenko}, {Kocsis}, {M{\'a}rka}, {M{\'a}rka}, {McKernan}, \&
  {O'Shaughnessy}}]{Yang2019b}
{Yang}, Y., {Bartos}, I., {Gayathri}, V., {et~al.} 2019,
  \href{http://dx.doi.org/10.1103/PhysRevLett.123.181101}{\prl},
  \href{https://ui.adsabs.harvard.edu/abs/2019PhRvL.123r1101Y}{123, 181101}

\bibitem[{{Yuan} \& {Narayan}(2014)}]{Yuan2014}
{Yuan}, F., \& {Narayan}, R. 2014,
  \href{http://dx.doi.org/10.1146/annurev-astro-082812-141003}{\araa},
  \href{https://ui.adsabs.harvard.edu/abs/2014ARA&A..52..529Y}{52, 529}

\bibitem[{{Yuan} {et~al.}(2018){Yuan}, {Yoon}, {Li}, {Gan}, {Ho}, \&
  {Guo}}]{Yuan2018}
{Yuan}, F., {Yoon}, D., {Li}, Y.-P., {et~al.} 2018,
  \href{http://dx.doi.org/10.3847/1538-4357/aab8f8}{\apj},
  \href{https://ui.adsabs.harvard.edu/abs/2018ApJ...857..121Y}{857, 121}

\bibitem[{{Zhu} {et~al.}(2021){Zhu}, {Zhang}, {Yu}, \& {Gao}}]{Zhu2021}
{Zhu}, J.-P., {Zhang}, B., {Yu}, Y.-W., \& {Gao}, H. 2021,
  \href{http://dx.doi.org/10.3847/2041-8213/abd412}{\apjl},
  \href{https://ui.adsabs.harvard.edu/abs/2021ApJ...906L..11Z}{906, L11}

\bibitem[{{Zrake} {et~al.}(2021){Zrake}, {Tiede}, {MacFadyen}, \&
  {Haiman}}]{Zrake2021}
{Zrake}, J., {Tiede}, C., {MacFadyen}, A., \& {Haiman}, Z. 2021,
  \href{http://dx.doi.org/10.3847/2041-8213/abdd1c}{\apjl},
  \href{https://ui.adsabs.harvard.edu/abs/2021ApJ...909L..13Z}{909, L13}

\end{thebibliography}
\bibliographystyle{aasjournalnolink}

\end{document}